\documentclass[aps,prl,onecolumn,superscriptaddress,groupedaddress]{revtex4}
\usepackage{graphicx}  % needed for figures
\usepackage{dcolumn}   % needed for some tables
\usepackage{bm}        % for math
\usepackage{amssymb}   % for math
\usepackage{slashed}
\usepackage{amsmath}
\usepackage{yfonts}
\usepackage{tikz}
\usepackage[utf8]{inputenc}
\usetikzlibrary{matrix,arrows,backgrounds,decorations.pathmorphing}
\usepackage{tikz-cd}
\usepackage[all]{xy}

\newcommand{\Bracket}[1]{\ensuremath{\left\langle#1\right\rangle}}

\begin{document}
\title{On SU(2) anomaly and Majorana fermions}
\author{Andrei Patrascu}
 \begin{abstract}
In this paper a loophole in the SU(2) gauge anomaly is presented. It is shown that using several topological tools a theory can be designed that implements the quantization of a single Weyl doublet anomaly free while keeping the non-abelian character of the particle in the theory.
This opens the perspective for non-Abelian statistics of deconfined particle like objects in 3+1 dimensions and for applications in Quantum Computing. Moreover, if this loophole cannot be closed, old 
arguments related to anomaly cancelations must be reviewed.
\end{abstract}
\affiliation{University College London, Department of Physics and Astronomy, London, WC1E 6BT, UK}
\pacs{04.20.Cv, 11.15.-q, 04.20.Fy, 03.70.+k}
\maketitle
\section{1. Introduction}
\par It is a fundamental feature of quantum mechanics that ordinary many particle systems in $3$ dimensions obey one of the two statistics: Bose-Einstein or Fermi-Dirac. Although in most of the textbook applications this fact is implemented in the form of a postulate, it can also be derived from topological arguments. The main advantage of the topological approach appears in the design of the topological quantum computers [44]. 
Following the ideas of reference [1] the indiscernability of particles can be implemented by means of restrictions imposed on the phase space. In fact, the symmetrization (or antisymmetrization) of the standard wavefunction can be traced back to the procedure of identifying the points in the phase space that 
differ by only a permutation $p \in S_{n}$ of the constituent particles. Here $S_{n}$ is the permutation group. Let $X^{N}$ be the $N$ dimensional phase space in the classical case. After identifying the points  that are equivalent with respect to $S_{n}$ we obtain the quotient space $X^{N}/S_{n}$. This space is locally isomorphic to $X^{N}$ but has a different topological structure. It also has several singular points where two or more particles occupy the same position. Its topological structure depends on the dimensionality of the original space. In $1$ or $2$ dimensions trajectories are infinitely connected and can be reduced to circles around singular points. In $3$ or more dimensions if one encircles a singular point once, the trajectory is homotopically equivalent to a circle. If one encircles the singularity twice, the resulting trajectory can be reduced to a single point without crossing the singularity. This particularity of the $3$ or higher dimensional spaces induces the possibility of $2$ distinct statistics: either Fermi-Dirac or Bose-Einstein. Moreover it eliminates the possibility for a point like particle to obey a non-abelian statistics precisely because of these topological particularities [2].
However, if one can add extra structure to the point-like particle this restriction may not apply.
This would imply the possibility of consistently defining non-abelian fractional statistics in $3+1$ dimensions between some special extended objects. Such objects (called hedgehogs) can trap Majorana zero-modes, a phenomenon of importance in the theory of high temperature superconductors (see for example the 2D p-wave superconductors as in ref. [41]-[43])
 The first proposal in this direction was formulated by Teo and Kane [3] who introduced hedgehogs of a $3$-component order parameter coupled to gapped fermionic excitations. These objects present a ``projective ribbon statistics'' [4] as far as multiple hedgehogs are associated to a non-local Hilbert space. Motions of the hedgehogs 
implement unitary transformations in the non-local Hilbert space. In this case exchanging identical particles leads to nontrivial unitary transformations of the quantum state (not simply a phase)[5]. This results in the hedgehogs obeying a non-abelian statistics. Moreover, hedgehog defects support real Majorana zero modes. It was an withstanding puzzle what happens to the Majorana zero modes when the relevant order parameter field begins to fluctuate. Also some researchers are still puzzled whether it is possible in principle to deconfine non-abelian particles in $3+1$ dimensions: if the order parameter field has nonzero stiffness, a single hedgehog is not a finite energy configuration. Although there will be finite energy hedgehog configurations (essentially with zero net hedgehog number), the confining force [6] between the hedgehogs will scale at least linearly with the distance between them. A way of avoiding this would be to gauge the rotation symmetry in the order parameter space[5]. Nevertheless a major obstacle in solving these puzzles is what is known as the $SU(2)$ gauge anomaly [7],[8]. In essence this anomaly states that in $3+1$ dimensions an $SU(2)$ gauge theory with the required fermion content i.e. a single Weyl doublet (or eight Majorana fields) cannot be defined consistently.
The previous attempts to solve this problem, by means of gauge anomaly cancelations, are well known [9]. The main idea was to eliminate anomalies by postulating new physical objects that may or may not have correspondence in reality. The impossibility of doing this and preserving the non-abelian structure at the same time, by using traditional methods has led to the belief that any theory presenting such an anomaly is unphysical. This paper does not make any general statements about what should be trusted more: currently available experiments or currently available mathematics. It just points out that in one case an alternative solution may be possible. The $SU(2)$ problem originates in the existence of a global anomaly (a gauge anomaly related to large gauge transformations i.e. transformations not simply connected to $Id$). The problem of non-abelian hedgehogs appeared mainly because of the assumption that, in order to maintain the non-abelian structure of one hedgehog, a full rotation of the hedgehog around a partner object must imply a change in the sign of the fermion parity. The general assumption was that this partner object must be a physical object. I hereby challenge precisely this assumption. The partner object should be merely a "theoretical measuring device" and can be constructed simply out of non-physical fields. I propose a theory where the non-abelian character is protected by a particular behavior of one hedgehog under rotations around a fictitious unphysical object introduced in the theory such that the theory itself is not otherwise altered. In fact this "object" appears due to a special global configuration of the functional space (in the sense of path integral quantization). I argue here that the $SU(2)$ anomaly problem can be eliminated by using some specific topological tools and some new ideas. 
Particularly, a global anomaly can be lifted if a suitable "measuring device" presenting a similar (compatible) global "anti-anomaly" is employed in the process of gauge fixing. This "measuring device" must be non-local in nature and is described either in terms of fields associated to the  BRST-dual-BRST quantization prescription or in terms of (co)homology with torsion coefficient groups (e.g. $\mathbb{Z}_{p}$, $p$-prime). It is important to remember that both the auxiliary fields and the coefficient groups in cohomology are arbitrary constructions that do not change the physical content of the theory. In the case of the BRST-dual-BRST quantization the integration over the artificial fields reconstructs the original theory. In the case of the interpretation using (co)homology with torsional coefficient groups, the universal coefficient theorem [36] tells us that the choice of coefficients is to a large extent arbitrary. One example of a situation where the coefficient groups in (co)homology are modified in order to obtain a "diluted" cohomology isomorphic with the Chech cohomology group with integer coefficients is [38]. 
These two ways of thinking (BRST-dual-BRST extension of a theory and the use of "exotic" coefficient groups in (co)homoloy) are to a large extent isomorphic. 
\par I start with a BRST-dual-BRST description, with the remark that, when introducing the effects of the here presented method on the Atiyah-Singer index theorem I will largely employ the (co)homological interpretation, making extensive use of torsion coefficient groups. 
\par The idea behind the BV-BRST approach is to generate a symplectic space suitable for geometric quantization. In general we start with a classical action $S[\cdot]$ depending on a set of fields. The classical theory provides the equations of motion via a minimization prescription. In a quantum theory however, we need extra information stored over the entire manifold associated to the fields. In order to access this information we complexify and exponentiate the classical action functional 
\begin{equation}
exp(iS[\cdot]):C\rightarrow A
\end{equation}
Here $C$ is the configuration space and $A$ is a resulting space. We then perform a functional integration over this construction. 
This definition is very formal. In practical cases the measure of the path integral is not always well defined. The configuration spaces are in general not even manifolds. Sometimes in order to obtain pertinent results a so called "cohomological integration" is necessary. When the theory we want to quantize has redundancies (also called gauge symmetries) there exist two possible approaches: when the gauge algebra is closed a BRST quantization procedure can be implemented. In general however, the gauge algebra is not closed. In this case an alternative method developed initially by Batalin and Vilkovisky is used. 

The algebra of the operators of the gauge symmetry can in general be defined as
\begin{equation}
 \frac{\delta^{l}R^{i}_{\alpha}}{\delta\phi^{j}}R_{\beta}^{j}-(-1)^{\epsilon_{\alpha}\epsilon_{\beta}}\frac{\delta^{l}R^{i}_{\beta}}{\delta\phi^{j}}R_{\alpha}^{j}=2R_{\gamma}^{i}T_{\alpha\beta}^{\gamma}(-1)^{\epsilon_{\alpha}}-4y_{j}E^{ji}_{\alpha\beta}(-1)^{\epsilon_{i}}(-1)^{\epsilon_{\alpha}}
\end{equation}
where $y_{j}=0$ represents the equation of motion, $E$ and $T$ represent coefficients, $R$ represent the (gauge) symmetry transformation operators and $\epsilon$ encodes the 
Grassmann parity of the associated field.
One can also define the BRST transformations of the original fields as $\delta\phi^{i}=R_{\alpha}^{i}[\phi]c^{\alpha}$ i.e. one can define the BRST symmetry transformations
via $R[\phi]$ and the associated ghost field $c^{\alpha}$ unambiguously. This is why, when no confusion is possible the terms 
$R^{i}_{\alpha}$, $R[\phi^{i},c,...]$ or the BRST transformation rule $\delta\phi^{A}=R^{A}[\phi^{B}]$ will be used alternatively as formal definitions. 

If $E=0$ the algebra is closed and the nilpotency of the BRST operator is naively verified. Imposing nilpotency on the fields $\phi^{i}$ we get
\begin{equation}
  0=\delta^{2}\phi^{i}=R_{\alpha}^{i}\delta c^{\alpha}+\frac{\delta^{l}R_{\alpha}^{i}c^{\alpha}}{\delta\phi^{j}}R_{\beta}^{j}c^{\beta}\\
\end{equation}
If we choose now 
\begin{equation}
 \delta c^{\gamma}=T^{\gamma}_{\alpha\beta}[\phi]c^{\beta}c^{\alpha}
\end{equation}
the nilpotency condition on the ``physical'' sector is satisfied and we obtain (considering $E=0$) 
\begin{equation}
 \frac{\delta^{l}R_{\alpha}^{i}c^{\alpha}}{\delta\phi^{j}}R_{\beta}^{j}c^{\beta}+R_{\gamma}^{i}T^{\gamma}_{\alpha\beta}c^{\beta}c^{\alpha}=0
\end{equation}
Also, using Jacobi identity one can easily show that $\delta^{2}c^{\gamma}=0$.
It will be seen later how this can be generalized for the case of BRST-anti-BRST transformations. 
If the algebra depends on the last term i.e. $E$ is not zero we have an open algebra and an non-nilpotent BRST transformation as acting on the initial fields.
This is a fundamental topological issue as the nilpotency of the $\delta$ operator (which is to be considered an exterior derivative in the BRST-cohomology) implies that $\delta^{2}=0$ and hence is the translation of the fact that all exact forms are closed. In other words, if $\delta$ is a boundary operator this non-nilpotency translates into the sentence "the boundary of a region has itself a boundary" which is impossible. 
The gauge fixed action constructed in the naive way would not be BRST invariant off-shell.
In order to solve this problem (i.e. to close the boundary of the region, if we insist on using the topological terminology) one has to introduce an artificial shift symmetry and to move the non-nilpotency from the transformation rules of the original fields to the transformation rules of the collective (and in a sense unphysical) fields [17], [18]. 
One certainly trivial way of enlarging the field space is by introducing two fields $A^{l}$ and $B^{l}$ such that
\begin{equation}
\begin{array}{l} 
\delta A^{l}=B^{l}\\
\delta B^{l}=0\\
\end{array}
\end{equation}
Obviously as the initial action does not depend on $A^{l}$ one can shift it with no practical effect. This shift would be a local symmetry and the fields $B^{l}$ would be the associated ghost-fields. It is precisely this idea that allows the redefinition of the field structure as will be seen further on. Having more fields of this type is of no physical consequence. What is important is the new perspective they can open upon the useful mathematical properties that can be added through them in the theory. For example it becomes possible to move undesirable aspects of the theory to the collective sector. It is also possible to transfer desirable properties to the physical field structure while using the unphysical sector in order to compensate the unphysical changes and to keep the same physical properties in the effective$^{1}$\footnotetext[1]{By effective it is usually understood a low energy, large scale equivalent of a theory obtained after the integration of microscopic details. However, it can also be considered to be a theory obtained after the integration of non-physical structures introduced only for the convenience of the calculation. In many situations those non-physical structures may reveal different ways of integrating in order to obtain an equivalent theory which is mathematically better defined} theory. In particular, if there are more symmetries available due to the extra fields, the interplay between them at the level of the BRST (-anti-BRST-dual-(anti)-BRST) transformations introduces additional freedoms that I am using in order to avoid the $SU(2)$ anomaly.
As one can see by now, the quantization prescription is not always trivial. One must specify what quantization means in the framework of path integrals. Essentially the special way in which the functional integration is performed assures the correct quantization of a classical theory. Moreover, the theory, defined by an action functional is by no means unique. It is well known that different representations can be chosen but in general in physics this amounts to the construction of effective low energy theories. However, this conclusion is not always necessary. By making different choices one can instead reveal useful properties in the theory that were not visible before.

The main source of the anomaly discussed here is related to a particularity of the $SU(2)$ group. In fact, its fourth homotopy group is non-trivial i.e. $\pi^{4}(SU(2))=Z_{2}$. This means that in order to reach identity with a gauge transformation one has to ``wrap'' two times around the whole $SU(2)$ group. If only one turn is performed no deformation to the identity is possible. After the second turn the identity is recovered but the two situations (with identity and without) are related in a continuous way by a gauge transformation. This means that the two regions are equally accessible via a gauge transformation and cannot be correctly distinguished. This problem of indiscernability is considered to be fundamental mainly because the fermion integration for a theory with $N$ massless Weyl fermion doublets may change sign under such a transformation[10]. It is clear at this moment that the two situations are related by a gauge transformation but the anomaly has its origin in the non-trivial global properties of the gauge group. In what follows I will alter the field-space such that the global non-triviality is being taken into account in a simple way. It is probably desirable to make a clarification at this point: it is not the special property of the $SU(2)$ group that is the problem here. The topological properties of the $SU(2)$ group are highly desirable and natural. The way we account for them however must change if we want to construct theories of this kind that also make sense. This can be accomplished either by changing the field structure and hence introducing artificial fields or, equivalently, by changing the coefficient groups in (co)homology [38], i.e. going to a torsion coefficient group. 
In order to be more specific let me start with a theory describing a single Weyl doublet:
\begin{equation}
 \int(d\psi d\bar{\psi})_{Weyl}exp(\bar{\psi}i\slashed{D}\psi)=\pm(det(i\slashed{D}))^{\frac{1}{2}}
\end{equation}
\par In this case the ambiguity of choosing the sign is essential. While picking an arbitrary sign for $(det(i\slashed{D}))^{1/2}$, in order to simultaneously satisfy the Schwinger Dyson equation one has to allow
a certain degree of freedom in the problem that will eventually change the sign of the square root without any control from our part. This aspect is not trivial as the path integral will gain an alternating sign which will amount in an ambiguity of the form ``$0/0$''. This problem can be related to the fact that the eigenvalues of the Dirac operator can be rearranged when a continuous gauge transformation is performed but only in such a way that an odd number of eigenvalues change sign from positive to negative. This of course generates a sign ambiguity. Nevertheless, one can introduce additional symmetry into the problem so that the Schwinger Dyson equation is satisfied in the form of a Ward Identity and the actual eigenvalues of the extended operator do not change the sign of the overall determinant. 
This can be done by keeping the same relevant information inside the theory [17], [18]. I underline that I eliminate the overall change in sign and not the relative change in sign between the hedgehog and the auxiliary structure to be introduced in the theory. I call this idea ``symmetry out of cohomology''.

\section{2. Preliminaries, artificial symmetries in gauge theories}
In this section I introduce, following mainly [18] and [19], a method of adding several independent gauge symmetries apart from the original gauge symmetry of the theory. At this moment only continuous gauge symmetries are considered. However, in the next sections and following [40] I will describe how a discrete symmetry can be added to the gauge structure. I also show here that it is possible to preserve the Schwinger-Dyson equations by means of the modified BRST algebra obtained in the process of adding auxiliary gauge symmetries. The fact that the Schwinger-Dyson equations are automatically fulfilled liberates us from the requirement of avoiding a benchmark that would notify us when we move from the branch of the $SU(2)$ group simply connected to $Id$ to the other branch. This possibility makes this preliminary section of major importance for the rest of this article. It also serves as a model calculation for the fact that integration of an extended theory can be done in several physically equivalent but mathematically different ways. Some integration prescriptions lead to significant simplifications.  
\par For now we can start with a pure Yang-Mills action $S[A_{\mu}]$. We call $A$ the connection which can be used to define a covariant derivative as 
\begin{equation}
D_{\mu}^{(A)}=\partial_{\mu}-[A_{\mu},\cdot]
\end{equation}
In order to enforce the Schwinger-Dyson equations as a result of the BRST algebra we may introduce a collective field 
\begin{equation}
A_{\mu}(x)\rightarrow A_{\mu}(x)-a_{\mu}(x)
\end{equation}
The transformed action $S[A_{\mu}-a_{\mu}]$ has two independent gauge symmetries. Due to the redundancies introduced by the collective field we can write the two symmetries in different ways. One way of doing it is 
\begin{equation}
\begin{array}{l}
\delta A_{\mu}(x)=\Theta_{\mu}(x)\\
\delta a_{\mu}(x)=\Theta (x)-D_{\mu}^{(A-a)}\epsilon(x)\\
\end{array}
\end{equation}
We may choose the original symmetry of the original field structure to be carried entirely by the collective field. The transformation of the original gauge field is always just a shift. $\Theta(x)$ includes arbitrary deformations. However, it only leaves the transformed field invariant. The action is also invariant under Yang-Mills gauge transformations of the transformed field itself. This is why two independent gauge transformations are being included. 
These two gauge symmetries have to be gauge fixed in the standard BRST fashion. We therefore introduce a suitable multiplet of ghosts and auxiliary fields. The shift symmetry of $A_{\mu}$ requires a vector ghost field $\psi_{\mu}(x)$. One Yang-Mills ghost field $c(x)$ will also be necessary. Gauge fixing the shift symmetry of $A_{\mu}$ by removing the collective field $a_{\mu}$ leads to the introduction of a corresponding antighost $A^{*}_{\mu}(x)$ and of an auxiliary field $b_{\mu}(x)$. 
\par The nilpotent BRST algebra now becomes 
\begin{equation}
\begin{array}{l}
\\
\delta A_{\mu}(x)=\psi_{\mu}(x)\\
\delta a_{\mu}(x)=\psi_{\mu}(x)-D_{\mu}^{(A-a)}c(x)\\
\delta c(x)=-\frac{1}{2}[c(x),c(x)]\\
\delta\psi_{\mu}(x)=0\\
\delta A^{*}_{\mu}(x)=b_{\mu}(x)\\
\delta b_{\mu}(x)=0\\
\\
\end{array}
\end{equation}
By adding 
\begin{equation}
-\delta[A^{*}_{\mu}(x)a^{\mu}(x)]=-b_{\mu}(x)a^{\mu}(x)-A_{\mu}^{*}(x)\{\psi^{\mu}-D_{(A-a)}^{\mu}c(x)\}
\end{equation}
to the Lagrangian we fix $a_{\mu}(x)$ to zero. At this point we can make the choice of integrating over pairs of ghosts and anti-ghosts. Hence we can integrate over $\psi_{\mu}(x)$ and $A^{*}_{\mu}(x)$ while keeping $c(x)$ unintegrated at this point. The extended but not yet fully gauge fixed action is
\begin{equation}
S_{ext}=S[A_{\mu}-a_{\mu}]-\int dx\{b_{\mu}(x)a_{\mu}(x)+A^{*}_{\mu}(x)[\psi^{\mu}(x)-D_{(A-a)}^{\mu}c(x)]\}
\end{equation}
with the partition function
\begin{equation}
Z=\int dA_{\mu}da_{\mu}d\psi_{\mu}dA^{*}_{\mu}db_{\mu}exp[\frac{i}{\hbar}S_{ext}]
\end{equation}
In order to continue, we first integrate out $a_{\mu}$ and $b_{\mu}$ and then, integration over $A^{*}_{\mu}$ leaves a trivial $\psi_{\mu}$ integral. In this way we obtain back the starting point, namely the Yang-Mills action $S[A_{\mu}]$ integrated over the original measure. 
\par We must insist that the Schwinger-Dyson equations involving the field $c(x)$ i.e. equations of the form 
\begin{equation}
0=\int dc \frac{\delta^{l}}{\delta c(x)}[F e^{\frac{i}{\hbar}[S]}]
\end{equation}
are satisfied automatically when employing the full, unbroken BRST algebra. In order to achieve this we have to introduce yet another collective field, say $\tilde{c}(x)$. We now shift the Yang-Mills ghost
\begin{equation}
c(x)\rightarrow c(x)-\tilde{c}(x)
\end{equation}
From this shift results a new fermionic gauge symmetry which we have to fix via the introduction of a new BRST ghost-antighost pair and an auxiliary field. We let the transformation of the new collective field $\tilde{c}(x)$ carry the BRST transformation of the original ghost. 
\begin{equation}
\begin{array}{l}
\\
\delta c(x)=C(x)\\
\delta \tilde{c}(x)=C(x)+\frac{1}{2}[c(x)-\tilde{c}(x),c(x)-\tilde{c}(x)]\\
\delta C(x)=0\\
\delta c^{*}(x)=B(x)\\
\delta B(x)=0\\
\\
\end{array}
\end{equation}
Now, in order to gauge fix $\tilde{c}(x)$ to zero we add the term
\begin{equation}
-\delta[c^{*}(x)\tilde{c}(x)]=B(x)\tilde{c}(x)-c^{*}(x)\{C(x)+\frac{1}{2}[c(x)-\tilde{c}(x),c(x)-\tilde{c}(x)]\}
\end{equation}
to the Lagrangian. This leads to the fully extended action 
\begin{equation}
S_{ext}=S[A_{\mu}-a_{\mu}]-\int dx\{b_{\mu}(x)a_{\mu}(x)+A_{\mu}^{*}(x)[\psi^{\mu}(x)-D^{\mu}_{(A-a)}\{c(x)-\tilde{c}(x)\}]-B(x)\tilde{c}(x)+c^{*}(x)(C(x)+\frac{1}{2}[c(x)-\tilde{c}(x),c(x)-\tilde{c}(x)])\}
\end{equation}
In the partition function all fields appearing above are being integrated except the field $c(x)$ for which another antighost $\bar{c}$ must still be introduced when the original Yang-Mills symmetry will be fixed eventually. 
The extended action and the functional measure is invariant under the following transformations
\begin{equation}
\\
\begin{array}{ll}
\delta A_{\mu}(x)=\psi_{\mu}(x), & \delta\psi_{\mu}(x)=0\\
\delta a_{\mu}(x) = \psi_{\mu}(x)-D_{\mu}^{(A-a)}[c(x)-\tilde{c}(x)], & \delta c(x)=C(x)\\
\delta A_{\mu}^{*}(x)=b_{\mu}(x), & \delta b_{\mu}(x)=0\\
\delta \tilde{c}(x)= C(x)+\frac{1}{2}[c(x)-\tilde{c}(x),c(x)-\tilde{c}(x)], & \delta C(x)=0\\
\delta c^{*}(x)=B(x), & \delta B(x)=0\\
\\
\end{array}
\end{equation}
The fields $A^{*}_{\mu}(x)$ and $c^{*}(x)$ are the antighosts of the collective fields which enforce the Schwinger-Dyson equations through shift symmetries. 
\par I used this preliminary chapter to show how additional shift symmetries can be used in order to encode the Schwinger-Dyson equations directly via the BRST algebra. The example given in this section is not new but serves as a model for the following chapters. It can be seen that by judiciously using artificial symmetries and gauge fixing, additional properties can be added to the original field structure. This is being done such that, by carefully integrating over the supplemental fields we obtain the same theory again. It will be clear in what follows that, by choosing to perform an extension of the field structure and a special field-integration, we can map an anomalous theory into another one carrying the same information in an effective way. This theory will not be plagued by the original anomaly.

\section{3. Theoretical approach}
 Let me start with a partition function plagued by the $SU(2)$ anomaly 
\begin{equation}
Z=\int d\psi d\bar\psi \int dA_{\mu}exp[-\int d^{4}x[(1/2g^{2})tr(F_{\mu\nu}^{2})+\bar\psi i\slashed{D}\psi]]
\end{equation}
where $A_{\mu}$ is the gauge field, $(1/2g^{2})tr(F_{\mu\nu}^{2})$ is the associated kinetic term and $F_{\mu\nu}$ is the field strength tensor (in other words, we have the connection $A$ and the curvature 2-form $F=dA\pm A\wedge A$). We also have $\bar\psi i\slashed{D}\psi$, the associated fermionic term.
I consider now the integration over the fermionic fields. This will present the problem related to the fermionic sign and the proposed solution of the $SU(2)$ anomaly. For the sake of brevity I will consider only the fermionic part. The kinetic term for the gauge fields is considered implicitly. The dynamics-less auxiliary fields to be used in this paper do not affect the kinetic term in a relevant way for this paper. However, they do affect the structure of the fermionic determinant in a way described in what follows. 
\par Let me start with showing how to introduce the Schwinger-Dyson equations as Ward identities[11]. Consider the actual form of the fermion field as
\begin{equation}
 \psi=\chi_{\alpha}
\end{equation}
where $\alpha$ represents the spin index. The covariant derivative is
\begin{equation}
 (D_{\mu}\chi)=\partial_{\mu}\chi-igA_{\mu \xi}T^{\xi}\chi
\end{equation}
Here $\xi$ is a gauge index.
Following the procedure by Batalin and Vilkovisky [16] I extend the example presented in the preliminary section and I insert now two auxiliary fields 
\begin{equation}
 A_{\mu \xi}\rightarrow A_{\mu \xi}-\phi_{1}-\phi_{2}
\end{equation}
These encode a trivial gauge symmetry representing a shift. The Jacobian associated to the above transformation is trivial. However, this symmetry involves additional freedoms to be employed in what follows. The new symmetry has to be gauge-fixed.
In doing so via the BRST-anti-BRST formalism (Becchi, Rouet, Stora and Tyutin [12]) 
the Schwinger-Dyson equation emerges as a Ward identity [17].
The field multiplets introduced are the ghosts $(\pi_{1},A_{2}^{*})$ and the antighosts $(A_{1}^{*},\pi_{2})$.
The BRST and anti-BRST transformations are as follow:
\begin{equation}
 \begin{array}{ll}
\delta_{1}A=\pi_{1} & \delta_{2}A=\pi_{2} \\
\delta_{1}\phi_{1}=\pi_{1}-A_{2}^{*} & \delta_{2}\phi_{1}=-A_{1}^{*} \\
\delta_{1}\phi_{2}=A_{2}^{*} & \delta_{2}\phi_{2}=\pi_{2}+A_{1}^{*} \\
\delta_{1}\pi_{1}=0 &  \delta_{2}\pi_{2}=0 \\
\delta_{1}A_{2}^{*}=0 & \delta_{2}A_{1}^{*}=0 \\
\end{array}
\end{equation}
Here $\delta_{1}$ and $\delta_{2}$ are respectively the BRST and anti-BRST transformations.
The next step is to impose gauge fixing. This is done in the standard way by adding more bosonic fields, call them $B$ and $\lambda$. The BRST transformation rules extend according to 
\begin{equation}
 \begin{array}{ll}
\delta_{1}\pi_{2}=B & \delta_{2}\pi_{1}=-B  \\
\delta_{1}B=0 & \delta_{2}B=0 \\
\delta_{1}A_{1}^{*}=\lambda-\frac{B}{2} & \delta_{2}A_{2}^{*}=-\lambda-\frac{B}{2} \\
\delta_{1}\lambda=0 & \delta_{2}\lambda=0 \\
\end{array}
\end{equation}
These rules imply the nilpotency conditions:
\begin{equation}
(\delta_{2}\delta_{1} + \delta_{1}\delta_{2})A=0
\end{equation}
\begin{equation}
(\delta_{2}\delta_{1} + \delta_{1}\delta_{2})\phi_{1}=0
\end{equation}
\begin{equation}
\delta_{1}^{2} = \delta_{2}^{2}=0
\end{equation}

One can chose the gauge fixing condition such that both auxiliary fields are fixed to zero by adding the BRST-anti-BRST closed term 
\begin{equation}
S_{col}=\frac{1}{2}\delta_{1}\delta_{2} [\phi_{1}^{2}-\phi_{2}^{2}]
\end{equation}
By using the BRST-anti-BRST transformations above this becomes
\begin{equation}
 S_{col}=-(\phi_{1}+\phi_{2})\lambda + \frac{B}{2}(\phi_{1}-\phi_{2})+(-1)^{a}A_{a}^{*}\pi_{a}
\end{equation}

which makes the gauge fixed action
\begin{equation}
S_{gf}=S_{0}[A-\phi_{+}]-\phi_{+}\lambda + \frac{B}{2}\phi_{-}+(-1)^{a}A_{a}^{*}\pi_{a}
\end{equation}
where $\phi_{\pm}=\phi_{1}\pm\phi_{2}$.
Here the index $a=1,2$ represents the field-antifield index and summation over it is implied.
Now the theory is well defined. 
At this moment the Schwinger-Dyson equation is encoded via an emerging Ward identity $\Bracket{\delta_{1}[A_{1}^{*}F[A^{\mu}]]} =0$ [17].
Alternatively this can be written as
\begin{equation}
 \begin{array}{l}
  0=<\delta_{1}[A_{\mu}^{*}F(A^{\mu})]>=\\
  \\
=\int d\mu [A_{\mu 1}^{*}\frac{\delta^{l}F}{\delta A_{\mu}}\pi_{1}+(\lambda-\frac{B}{2})F(A_{\mu})]e^{\frac{i}{\hbar}S_{gf}}\\
 \end{array}
\end{equation}
Here $F$ is a general functional on the fields $A_{\mu}$ and $\frac{\delta^{l}}{\delta A_{\mu}}$ is the left functional derivative. It gains a sign with respect to the right derivative when acting on fermionic
fields.

The next step is to implement an artificial discrete symmetry that corresponds
to the anti-unitary time reversal. Indeed this can be done if one considers the de-Rham cohomology. Here we cannot restrict ourselves to the BRST-anti-BRST operators but we need all the operators of the de-Rham cohomology, including the dual-BRST-anti-BRST operators. 
One has to acknowledge that the field structure of a theory is not fundamental. In fact, following [18] the antifields used to extend the usual field structure can be considered on equal footing with the usual fields. In most of the cases however one starts from a theory where these have been integrated out in advance. This however, is not necessary and implies some information loss when thinking at the procedure of integrating out the auxiliary fields. If one starts with a theory containing fields and antifields and integrates them in a symmetric way the resulting theory can be constructed such that its field space has a Kahlerian structure. Imposing a Kahlerian structure is not an ad-hoc construction. First, the complexification required implies a simpler encoding of non-local properties and second, the holomorphicity of the field space will play the role of a "benchmark" between the region simply connected to $Id$ in $SU(2)$ and the other region.
The theory constructed in this way also manifests a discrete symmetry. The Kahler structure makes this discrete symmetry appear in the form of an anti-unitary time reversal symmetry induced by the Hodge star operator. 
 This "mirror" symmetry cannot introduce divergencies in the theory. However, the Kahler-structure imposed over the field space, which can be interpreted as a choice of a Kahler quantum polarization, assures that the extra fields protect on one side the non-abelian statistics of the remaining hedgehog but also assure a constant overall sign in the full fermionic determinant.

In order to be more specific (for more details see appendix B) if we are given a differential manifold $M$ and a tensor of type $(1,1)$ $J$ such that $\forall p \in M$, $J_{p}^{2}=-1$, the tensor $J$ will give a structure to $M$ with the property
that the eigenvalues of it will be of the form $\pm i$. This means that $J_{p}$ is an even dimensional matrix and $M$ is an even manifold. It also follows that $J_{p}$ can divide a complexified space at a point $p$
in two disjoint vector subspaces 
\begin{equation}
 T_{p}M^{C}=T_{p}M^{+}\oplus T_{p}M^{-}
\end{equation}
\begin{equation}
 T_{p}M^{\pm}=\{Z\in T_{p}M^{C}\mid J_{p}Z=\pm iZ\}
\end{equation}
One can also introduce two projection operators
\begin{equation}
 P^{\pm}:T_{p}M^{C}\rightarrow T_{p}M^{\pm}
\end{equation}
\begin{equation}
 P^{\pm}=\frac{1}{2}(1\pm iJ_{p})
\end{equation}
which will decompose $Z$ as $Z=Z^{+}+Z^{-}$. This construction will generate a holomorphic and an antiholomorphic sector: $Z^{\pm}=P^{\pm}Z\in T_{p}M^{\pm}$, $T_{p}M^{+}$ being the holomorphic sector.
A complex manifold appears when demanding that given two intersecting charts $(U_{i},\gamma_{i})$ and $(U_{j},\gamma_{j})$, the map $\psi_{ij}=\gamma_{j}\gamma_{i}^{-1}$ from $\gamma_{i}(U_{i}\cap U_{j})$
 to $\gamma_{j}(U_{i}\cap U_{j})$ is holomorphic. Here $\gamma_{i}$ and $\gamma_{j}$ are chart homeomorphisms and $\psi_{ij}$ is the transition map. In this case the complex structure is given independently from the chart by 
\begin{equation}
 J_{p} = \left( \begin{array}{cc} 0 & 1 \\ -1 & 0 \end{array} \right) \forall p\in M
\end{equation}
In the complex case there is a unique chart-independent decomposition in holomorphic and antiholomorphic parts. This means we can now choose as a local basis for those subspaces the vector
$(\frac{\delta}{\delta z^{\mu}},\frac{\delta}{\delta\bar{z}^{\mu}})$ where ($z^{\mu},\bar{z}^{\mu}$) are the complex coordinates
such that the complex structure becomes
\begin{equation}
 J_{p} = \left( \begin{array}{cc} i1 & 0 \\ 0 & -i1 \end{array} \right) \forall p\in M
\end{equation}

The additional structure over the field space can be introduced in several different ways. Here, I show a method suitable for the 
current problem. 
Consider another extension of the field structure in the following way:
\begin{equation}
 \{A_{\mu}\}\rightarrow \{A_{\mu},A^{\Omega},\bar{A}^{\dot{\Omega}}\}
\end{equation}
This method is similar with the previous method of introducing additional auxiliary (unphysical) fields. However, the way these fields are introduced here is special because they also carry topological information. They are also introduced in such a way that a Kahler structure emerges over the resulting field space. This is also ensured by the special form of the matrix $h$ introduced in what follows
\begin{equation}
\begin{array}{c}
 Z=\int (d\psi d\bar{\psi} dA_{\mu} dA^{\Omega} d\bar{A}^{\dot{\Omega}} d\mu) \\
 \\
exp[\bar{\psi}(\bar{\sigma}^{\mu}_{\alpha\beta}(i\partial_{\mu}+g(A_{\mu}-\phi_{+}+ig_{\mu\nu}A^{\Omega}h_{\Omega\dot{\Omega}}^{\nu}\bar{A}^{\dot{\Omega}})))\psi- \\
\\
- \phi_{+}\lambda+\frac{B}{2}\phi_{-}+(-1)^{a}A_{a}^{*}\pi_{a}]\\
\end{array}
\end{equation}
where $d\mu$ represents the integration measure with respect to the rest of the auxiliary fields, $\bar{\sigma}=(1,-\overrightarrow{\sigma})$ and $g_{\mu\nu}$ is the standard space-time metric.
The matrix $h_{\Omega\dot{\Omega}}$ assures in this case that the gauge fixing procedure is done in a BRST-anti-BRST invariant way. It also assures that entries between the Grassmann odd and Grassmann even sectors vanish. This will imply that a term of the form $\phi_{A}h^{AB}\phi_{B}$ has ghost-number zero and even Grassmann parity.
Otherwise $h^{AB}$ has a flexible form required in defining a corresponding metric over the field space. The indexes $\Omega$ and $\dot{\Omega}$ refer to an internal space used to define the Kahler structure over the field space.

Of course gauge fixing is needed. In order to do this one may add the closed form
\begin{equation}
 \Omega=i\delta\bar{\delta}(K(A,\bar{A})-ih_{\Omega\dot{\Omega}}A^{\Omega}\bar{A}^{\dot{\Omega}})
\end{equation}
Here $\Omega$ plays the role of the Kahler form. $K(A,\bar{A})$ is the Kahler potential and it has the property of generating the metric when the co-exterior and anti-co-exterior derivatives act on it
\begin{equation}
\begin{array}{l}
 K_{i}:U_{i}\rightarrow R\\
 \\
G=\delta\bar{\delta}K_{i}\\

\end{array}
\end{equation}
where $\delta$ and $\bar{\delta}$ can be associated to the dual-BRST and dual-anti-BRST operators and $G$ is the induced metric over the field space.
This will ensure the Kahler structure and will not alter the rest of the structure as it is a closed form under the co-exterior derivative
\begin{equation}
 d^{*}\Omega=(\delta+\bar{\delta})\Omega = 0
\end{equation}
It also forms a class with respect to the de-Rham cohomology and it is not an exact form i.e. it cannot be written as $\Omega=d^{*}\Psi$. 
The fact that this form is closed but not exact is implied by the fact that the construction of the field space (a compact manifold in this case) was designed such that the Kahler form was made manifest and a Kahler form cannot be exact on a compact space. When the operators of the direct and dual sector are made manifest the theory is best described by the de-Rham cohomology and the Kahler form represents a distinct class in this cohomology. 

\par We now define the Hodge star operator in the following way (see appendix B): let $\alpha$ and $\beta$ be two N-forms, 
\begin{equation}
\alpha,\beta \in \bigwedge^{N}
\end{equation}
then $*\beta$ is defined such that given the metric $G$ over the considered manifold and $dv_{G}$ a unit N-vector we have
\begin{equation}
 \alpha\wedge *\beta = G_{p}(\alpha,\beta)dv_{G}
\end{equation}
$(**)=1$ on $\bigwedge^{N}$ (we are at the level of the first doubling so, already $N=2d$) which means that $\bigwedge^{N}$ splits into eigenspaces as 
\begin{equation}
 \bigwedge^{N}=\bigwedge_{+}^{N}+\bigwedge_{-}^{N}
\end{equation}
where the two eigenspaces correspond to eigenvalues +1 and -1 respectively. A N-form which belongs to $\bigwedge_{+}^{N}$ is called self-dual whereas if it belongs to the other eigenspace it is called 
anti-self-dual. An important remark to be done here is that given a p-vector $\lambda\in\bigwedge^{p}V$ then $\forall \theta\in\bigwedge^{n-p}V$ there exists the wedge product such that $\lambda\wedge\theta\in\bigwedge^{n}$.
The (anti)BRST and dual-(anti)BRST operators are then equivalent to the operators:
\begin{equation}
 d:\bigwedge^{k}\rightarrow \bigwedge^{k+1}
\end{equation}
\begin{equation}
 d^{*}=*d*:\bigwedge^{k}\rightarrow \bigwedge^{k-1}
\end{equation}
\begin{equation}
 \Delta=d d^{*}+d^{*} d:\bigwedge^{k}\rightarrow \bigwedge^{k}
\end{equation}
In the context of algebraic geometry these are in order: the exterior differential, the coexterior(dual) differential and the Laplace operator (see appendix B). 
The exact and co-exact forms are orthogonal. 
Here we have the exterior derivative 
\begin{equation}
d=\delta_{1}+\delta_{2}
\end{equation}
and its dual
\begin{equation}
d^{*}=*d*=\delta+\bar{\delta}
\end{equation}

This gives rise to a suitable candidate for a ``barrier'' that would allow one to discern  whether one is on the side connected to $id$ or on the other side or, probably better formulated, it would make the two parts properly separated with $id \leftrightarrow id*$.
The integration is performed in the same way with the exception that due to the Kahler structure any change in the sign in one sector is compensated by a corresponding change in the dual sector. This is being done while allowing the fermionic parity of an individual hedgehog to vary when considering its behavior under relative rotations around the fictive Kahler structure associated to it. Please note how dual-space gauge fixing and the implementation of a Kahler structure interplay in order to keep an overall positive fermionic determinant and a non-abelian statistics for the hedgehog. Of course this would not be possible if additional structure could not be added to the problem in a chomologically invariant way i.e. for fundamental particles. Fortunately the condensed matter background allows this auxiliary and otherwise inert structure in the theory. 
\par In a more illustrative way, one could imagine that in every point in the space considered, one could add a circular space. While the integration would diverge due to uncontrollable integration over the circular space at each point, the gauge fixing would solve this problem by picking a single representative in the circular space. However, the choice of a representative in the internal circular space would not solve on its own the change in sign due to the topology of the original space. The solution in this case is the addition of a dual space to this construction. In this case a dual circular space will also introduce an uncontrollable integration and it will also have to be gauge fixed. However, this can be done in the functional space such that the global change in sign is compensated. In fact, a discrete symmetry is constructed in the action functional from the way in which the gauge is fixed over the direct and dual spaces. If the resulting field-space is Kahlerian (as intended in this paper) the discrete symmetry, induced by the Hodge star operator, mimics the time reversal symmetry and conserves an apparent non-abelian statistics. It is this discrete artificial symmetry that plays here the role of a non-local measuring device, to be encoded later in the coefficient groups in cohomology. As for now, this $Z_{2}$ symmetry has the role of controlling the change in sign during the integration behaving as an artifact of the mathematical procedure. 
While associated auxiliary fields are used in order to extend the field space via the BV-BRST method, the final construction of a discrete internal symmetry (in this case a $Z_{2}$ symmetry) does not lead to an additional associated gauge field. What is important is the process of covering the manifold with overlapping patches, each equipped with its own set of conventions. It is topologically interesting to know if the local conventions can be patched together such that they induce a global convention. On simply connected manifolds a global convention is always possible. On non-simply connected manifolds the situation becomes somehow more complicated [40]. There, one has to make a choice related to what structure to impose over the manifold such that the induced patching conventions make sense. In general this is not possible (see for example [40] but also any textbook on Cech cohomology). For the current issue however, if we can add a Kahler structure over the field space such that a $Z_{2}$ time reversal type symmetry emerges, the global inconsistency can be taken into account by using the preservation of (anti-)holomorphic vectors on Kahler manifolds. This can be translated in terms of cohomology with coefficients in torsion groups i.e. a non-trivial cohomology with coefficients in $Z$ may translate into a trivial cohomology when analyzed via coefficients in a torsion group (see [34], [36]).

\par This method allows a redefinition of the Dirac operator so that the problem becomes well defined. The first investigation of the behavior of the Dirac operator as a function of the metric is due to Hitchin [39]. The complexification that appears as a part of the construction of the Kahler space is important mainly because we wish to deal with the global (topological) properties of the $SU(2)$ group and we have to model those in an appropriate way.

In fact one observes that the integral changes to
\begin{equation}
\begin{array}{c}
\int(d\psi d\bar{\psi})_{Weyl-Aux} exp[\bar{\psi} (i\slashed{D}^{(A^{\mu}-\phi_{+})}-A^{\Omega}h_{\Omega\dot{\Omega}}\bar{A}^{\dot{\Omega}})\psi]\\
\\
=det(i\slashed{D}^{(A^{\mu}-\phi_{+})}-A^{\Omega}h_{\Omega\dot{\Omega}}\bar{A}^{\dot{\Omega}})\\
\end{array}
\end{equation}
and by adding the Kahler term in the partition function one has 
\begin{equation}
\begin{array}{c}
Z=\int (dA_{\mu} dA^{\Omega}d\bar{A}^{\dot{\Omega}}d\mu)det(i\slashed{D}^{(A^{\mu}-\phi_{+})}-A^{\Omega}h_{\Omega\dot{\Omega}}\bar{A}^{\dot{\Omega}})\\
\\
exp[i\delta\bar{\delta}(K(A,\bar{A})-ih_{\Omega\dot{\Omega}}A^{\Omega}\bar{A}^{\dot{\Omega}})-\phi_{+}\lambda+\frac{B}{2}\phi_{-}+(-1)^{a}A_{a}^{*}\pi_{a}]\\
\end{array}
\end{equation}
but now the determinant is defined for a Kahler structure. A gauge transformation in the Kahler space will have the form
\begin{equation}
A\longrightarrow A+i(\Lambda-\Lambda^{\dagger})
\end{equation}
so, while preserving this form a transformation of this kind will not alter the structure of the theory. This imposes a specific form of the $h_{\Omega\dot{\Omega}}$ tensor.  Gauge fixing is performed by 
a suitable choice of the introduced functions. Of course, all the fields can be eliminated  by integrating out pairs of fields and antifields or by integrating only one member of the pair but suitably changing the transformation rules of the remaining fields.
It is extremely important to notice that the integration is performed over both the fields $A^{\Omega}$ and $A^{\dot{\Omega}}$ and that they cover both the direct and the dual regions. Because of the Kahler structure constructed over the field space, they behave in the desired way i.e. they prevent the change in sign. Please note that the geometry of the field space becomes Kahlerian.
The direct and dual field sectors combine giving in any case a positive determinant. 
\par It may be instructive to repeat here the possible definitions of a Kahler manifolds. Indeed one can say equivalently that a Kahler manifold is 
\begin{itemize}
\item a symplectic manifold $(K,\omega)$ with an integrable almost complex structure
\item a Hermitian manifold with the associated Hermitian form closed
\item a manifold where the complex structure, the Riemannian structure and the symplectic structure are mutually compatible
\end{itemize}
Indeed, due to these definitions, on a Riemannian manifold $M$ it is always possible to choose Riemannian normal coordinates at any point $p\in M$. In fact, these coordinates are those in which the metric takes its canonical form $g_{ab}=\delta_{ab}$ at the point $p$ and all its first derivatives vanish at that point. If we consider a general Hermitian manifold, the holomorphicity of the coordinates for which this is true is not always assured. In order to keep the holomorphicity and the canonical form for the metric at a point we need to have Kahler manifolds. As a simple situation, consider the Levi-Civita connection and its Christoffel symbols. A Kahler manifold assures us that if we define (anti-)holomorphic vectors at a point, their parallel transport is also into (anti-)holomorphic vectors. Otherwise stated, in terms of the Christoffel symbols, $\Gamma_{jk}^{i}$ or $\Gamma_{\bar{j}\bar{k}}^{\bar{i}}$ may be non-zero but all mixed symbols like $\Gamma_{jk}^{\bar{i}}$ must be identical to zero. This also means that $n$-dimensional Kahler manifolds are by definition $2n$-dimensional Riemannian manifolds with holonomy group contained in $U(n)$. From a physical point of view, this special property of the field space assures us that by performing a "large" gauge transformation we do not mix holomorphic vectors with anti-holomorphic vectors. This property is crucial because it is only by this that the integration on each branch of the new field-space can be performed meaningfully and it is only by this that the full topological information available initially only when considering both of the branches now becomes available only on one branch.

Moreover, the whole construction presented until here, i.e. the introduction of a symplectic structure due to the use of the BV-BRST construction, the complexification and the special choice of the complex structure, such that integrability is insured, all combine in order to have the Kahler property for the field space. 
\par One may ask if this would have been possible using fewer properties. It can naively be argued that a more parsimonious employment of algebraic structures might suffice. This, unfortunately is not the case. Giving up the complex structure would make it impossible to encode the global information in a suitable way. Giving up the symplectic structure would make the physics meaningless, not to say that the formal quantization prescription would loose much of its relevance. Finally, keeping only the three structures, hence a Kahler manifold, assures that all the information can be mapped in the branch over which the integration is meaningful.

\par All this will have some consequences on the interpretation of the Atyiah-Singer index theorem [14] and the flux, both essential ingredients in the formulation of the SU(2) anomaly [7]. 
First, the field structure is "supplemented" by the Kahler condition such that the eigenvalues do not change the sign. The method of extending the field structure plays the role of a ``topological regularizer''. As one homotopically changes the path in the gauge space the change in sign from one part of the Kahler sector is compensated by the other part. Moreover, due to a property of the Kahler structure (i.e. local holomorphicity is preserved while 
performing a gauge transformation) this compensating property will be preserved over the entire gauge group. 
Second, a specific choice of the Kahler potential in the functional described above will not affect the physical content of the theory but will modify the set of symmetries of the problem in the desired way. 
\par It is important to explain in what sense the physical content of the theory remains unaffected. Indeed, in the most general case, when additional fields are introduced the content of the theory changes dramatically [37]. In the present case, the additional fields are introduced such that they become relevant in a global, topological sense. They are constructed such that they compensate for the global anomaly only. One may say that they belong to the same cohomology class as the physical sector but this statement is too weak. The best way of explaining this situation is to remark that the internal, circular spaces introduced in the theory can be described from an algebraic standpoint as periodic coefficient groups in cohomology. These generate a torsion visible globally that alters the classes in the cohomology group by merging some of them and separating others. This effect is of no direct physical relevance due to what is known as the universal coefficient theorem [34], [36]. However, in order to make sure that we can use any coefficient group we want, we must take the, now modified $Ext$ and/or $Tor$ groups correctly into account in the universal coefficient theorem [33]. This will be the subject of the last section in the context of the Atiyah-Singer index theorem. It is also explained in more detail in section $7$, subsection $C$ of [34]. 
In the most general sense, an anomaly results from the fact that a certain mathematical description, suitable for a specific context becomes unsuitable for a different context. What one has to do is to lift the original mathematical description to the other context making all the changes that are necessary. For example, a functional $a=a(A,\omega)$ where $A$ is the gauge potential and $\omega$ is a ghost field, is called a "true" anomaly if it satisfies the Wess-Zumino consistency condition, $\delta a=0$ but there is no local functional $\Lambda_{loc}(A)$, such that a redefinition of the effective action $\Gamma$ as $\Gamma\rightarrow \Gamma + \Lambda_{loc}$ would cancel the anomaly itself. However, there exist other changes, not visible at the level of perturbative calculations, that can eliminate the anomalous situation. These changes are given by homological algebra, for example by a specific choice of coefficient groups in cohomology, their effects being undetectable locally. Of course, a change in the original theory must occur, as the original theory was not well defined in the new context. However, these changes preserve the topological properties, in this case of the $SU(2)$ group. Universal coefficient theorems will tell us where precisely the missing information is stored. The advantage of the new constructions is that they are more suitable for the new case.

\section{4. Internal spaces and duality}
In the construction of the extended field space I used an internal space in order to naturally define duality. To be more explicit, I will follow here reference [26] to show that the construction of an internal space is useful in this context and that a discrete $Z_{2}$ symmetry can appear. 
I start by following reference [26] with an example of even dimensional $(2n)$ 
electrodynamics. Let $A$ be a general $(n-1)$ form and $F_{k_{1},...k_{n}}$ its associeated field
strength:
\begin{equation}
 F_{k_{1}...k_{n}}=\partial_{ [ k_{n}}A_{k_{1}...k_{n-1} ] }
\end{equation}
\begin{equation}
 *F^{k_{1}...k_{n}}=\frac{1}{n!}\epsilon^{k_{1}...k_{2n}}F_{k_{n+1}...k_{2n}}
\end{equation}
Given the action, the equation of motion and the Bianchi identity as
\begin{equation}
 S=-c_{n}\int d^{2n}x F_{k_{1}...k_{n}}F^{k_{1}...k_{n}}
\end{equation}
\begin{equation}
 \partial_{k_{1}}F^{k_{1}...k_{n}}=0
\end{equation}
\begin{equation}
 \partial_{k_{1}}*F^{k_{1}...k_{n}}=0
\end{equation}
($c_{n}$ is a constant, $k_{j}$ is the tensorial index) we can see that at the level of the Bianchi identity and the equation of motion the dual operation is a symmetry.
Nevertheless, in general the second power of the dual operation has a different structure depending on the dimension of the space:
\begin{equation}
**F=\left\{ \begin{array}{rcl} F & if & D=4k-2\\-F & if & D=4k \end{array}\right.
\end{equation}
As one can see the dual * is not well defined for the 2-dimensional $(2D)$ scalar or for the 4k-2 dimensional extensions. Its definition has been enlarged [26]
by making an internal structure of the potentials in the theory manifest. 
One should note that this has been achieved by using a canonical transformation and that the same can be achieved via BRST. I will enlarge the set of fields 
(alternatively the Hilbert space) by
giving them an internal structure of the form $(\alpha,\beta)$.
The dual operation is now defined as
\begin{equation}
\begin{array}{ll}
 \tilde{F}^{\alpha}=\epsilon^{\alpha\beta}*F^{\beta},& D=4k
\end{array}
\end{equation}
\begin{equation}
\begin{array}{ll}
\tilde{F}^{\alpha}=\sigma_{1}^{\alpha\beta}*F^{\beta},& D=4k-2
\end{array}
\end{equation}
\begin{equation}
 \tilde{\tilde{F}}=F
\end{equation}
$\sigma_{1}^{\alpha\beta}$ being the first Pauli matrix. In this case self and anti self dualities are well defined in any $D=2k$ dimensional space. 
One can start with the first order form of the theory:
\begin{equation}
 S=\int d^{D}x[\Pi\cdot\dot{A}-\frac{1}{2}\Pi\cdot\Pi-\frac{1}{2}B\cdot B + A_{0}(\partial\cdot\Pi)]
\end{equation}
Maxwell's Gauss constraint can be generalized to be precisely the extended curl $(\epsilon\partial)=\epsilon_{k_{1}k_{2}...k_{D-1}}\partial_{k_{D-1}}$. Then
\begin{equation}
 \Pi=(\epsilon\partial)\cdot\phi
\end{equation}
\begin{equation}
 B=(\epsilon\partial)\cdot A
\end{equation}
where $\phi$ is a $(\frac{d}{2}-1)$-form potential, $A$ is a generalization of the vector potential, $A_{0}$ is the general multiplier that enforces the Gauss
 constraint, the antisymmetrization of $\partial$ is defined as
\begin{equation}
 (\epsilon\partial)=\epsilon_{k_{1}k_{2}...k_{D-1}}\partial_{k_{D-1}}
\end{equation}
and in general the notation 
\begin{equation}
  \Phi\cdot\Psi=\Phi_{[k_{1}...k_{D-1}]}\Psi_{[k_{1}...k_{D-1}]}
\end{equation}
is used to imply antisymmetrization via the brackets.
Now I construct an internal space of potentials where duality symmetry is manifest ($\Phi^{+}$ and $\Phi^{-}$ represent the new field structure). The dual 
projection can be defined now as a canonical transformation of the fields in the following way:
\begin{equation}
 A=(\Phi^{+}+\Phi^{-})
\end{equation}
\begin{equation}
\Pi=\eta(\epsilon\partial)(\Phi^{(+)}-\Phi^{(-)})
\end{equation}
\begin{equation}
 \eta=\pm1
\end{equation}
The action can be rewritten in terms of these fields as
$$
 S=\int d^{D}x\{\eta[\dot{\Phi}^{(\alpha)}\sigma_{3}^{\alpha\beta}B^{(\beta)}+\dot{\Phi}^{(\alpha)}\epsilon^{\alpha\beta}B^{(\beta)}]-B^{(\beta)}\cdot B^{(\beta)}
$$
where $B^{(\beta)}=(\epsilon\partial\cdot\Phi^{(\beta)})$ and $\sigma_{3}^{(\alpha\beta)}$ and $\sigma_{2}^{(\alpha\beta)}=i\epsilon^{(\alpha\beta)}$ are the Pauli marices.
We see that the symplectic part factorizes in two parts: one involving the third Pauli matrix and the other one the second Pauli matrix. 
For a dimension $D=4k$ the first term is the generalization of the $2D$ chiral bosons. The $Z_{2}$ symmetry manifests itself in the 
transformation $\Phi^{(\pm)} \longleftrightarrow \Phi^{(\mp)}$. The second
term becomes a total derivative. For $D=2K$ the first term becomes a total derivative and the second term explicitly shows the symmetry of SO(2).
Although the complete diagonalization of the action in 3D cannot be done in coordinate space a dual projection is possible in the momentum space [26].
Let me introduce a two-basis $\{\hat{e}_{a}(k,x),$ $a=1,2\}$ with $(k,x)$ being conjugate variables and the orthonormalization condition given as
\begin{equation}
 \int dx \hat{e}_{a}(k,x)\hat{e}_{b}(k',x)=\delta_{ab}\delta(k,k')
\end{equation}
The vectors in the basis can be chosen to be eigenvectors of the Laplacian, $\nabla^{2}=\partial \partial$ and
\begin{equation}
 \nabla^{2}\hat{e}_{a}(k,x)=-\omega^{2}(k)\hat{e}_{a}(k,x)
\end{equation}
The action of $\partial$ over the $\hat{e}_{a}(k,x)$ basis is
\begin{equation}
 \partial\hat{e}_{a}(k,x)=\omega(k)M_{ab}\hat{e}_{b}(k,x)
\end{equation}
The two previous equations give
\begin{equation}
 \tilde{M}M=-I
\end{equation}
where $\tilde{M}_{ab}=M_{ba}$.
The canonical scalar and its conjugate momentum have the following expansion
\begin{equation}
 \Phi(x)=\int dk q_{a}(k)\hat{e}_{a}(k,x)
\end{equation}
\begin{equation}
 \Pi(x)=\int dk p_{a}(k)\hat{e}_{a}(k,x)
\end{equation}
where $q_{a}$ and $p_{a}$ are the expansion coefficients. 
The action appears in this representation as a two dimensional oscillator. The phase space is now four dimensional, representing two degrees of freedom per mode,
\begin{equation}
 S=\int dk\{p_{a}\dot{q}_{a}-\frac{1}{2}p_{a}p_{a}-\frac{\omega^{2}}{2}q_{a}q_{a}\}
\end{equation}
now we can introduce the following canonical transformation
\begin{equation}
 p_{a}(k)=\omega(k)\epsilon_{ab}(\varphi_{b}^{(+)}-\varphi_{b}^{(-)})
\end{equation}
\begin{equation}
 q_{a}(k)=(\varphi_{a}^{(+)}+\varphi_{a}^{(-)})
\end{equation}
The action becomes $S=S_{+}+S_{-}$ where
\begin{equation}
 S_{\pm}=\int dk\omega(k)(\pm\dot{q}_{a}\epsilon_{ab}q_{b}-\omega(k)q_{a}q_{a})
\end{equation}
As expected, this action presents the $Z_{2}$ symmetry under the transformation $\varphi_{a}^{\alpha}\rightarrow \sigma_{1}^{\alpha\beta}\varphi_{a}^{\beta}$.

This is a particular example. However, the field-antifield prescription used in the main 
paper has practically a similar role and is defined in general. It generates a symplectic even dimensional field space suitable for quantization. It also defines an analogues for the Hodge-* operators.

\section{5. Hodge star as a discrete symmetry}
For an example of how the Hodge star induces a discrete symmetry I follow ref. [27]-[29].
The main idea there was to represent the Hodge decomposition operators $(d,\delta,\Delta)$  as some symmetries of a given BRST invariant Lagrangean of a gauge theory. 
In general, the Hodge decomposition theorem (see section 5) states that on a compact manifold any $n$-form $f_{n}(n=0,1,2,...)$ can be uniquely represented as the sum of a harmonic form $h_{n} 
(\Delta h_{n}=0,dh_{n}=0, \delta h_{n}=0)$, an exact form $de_{n-1}$ and a co-exact form $\delta c_{n+1}$ as
\begin{equation}
 f_{n}=h_{n}+de_{n+1}+\delta c_{n+1}
\end{equation}
where here $d$ is the exterior derivative, $\delta$ is its dual and $\Delta$ is the Laplacian operator $\Delta=d\delta+\delta d$.
In order to identify the dual BRST transformation, one has to observe that while the direct BRST transformations leave the two form $F=dA$ in the construction of a 
gauge theory invariant and transform the Dirac fields like a local gauge transformation, the dual-BRST transformations leave the previous gauge fixing term invariant
 and transform the 
Dirac fields like a chiral transformation.
So, as a practical example, I can start like the authors of [29] from a BRST invariant lagrangean for QED noting that generalizations for non-abelian gauge theories with 
interactions exist in the literature as well. 
\begin{equation}
 L_{B}=-\frac{1}{4}F^{\mu\nu}F_{\mu\nu}+\bar{\psi}(i\gamma^{\mu}\partial_{\mu}-m)\psi-e\bar{\psi}\gamma^{\mu}A_{\mu}\psi+B(\partial A)+\frac{1}{2}B^{2}-i\partial_{\mu}\bar{C}\partial^{\mu}C
\end{equation}
$F^{\mu\nu}$ being the field strength tensor, $B$ is the Nakanishi-Lautrup auxiliary field and $C$, $\bar{C}$ are the anticommuting ghosts.
The BRST transformations that leave this Lagrangian invariant are
\begin{equation}
 \begin{array}{ll}
  \delta_{B}A_{\mu}=\eta\partial_{\mu}C & \delta_{B}\psi=-i\eta e C \psi\\
\delta_{B}C=0 & \delta_{B}\bar{C}=i\eta B\\
\delta_{B}\bar{\psi}=i\eta e C\bar{\psi} & \delta_{B}F_{\mu\nu}=0\\
\delta_{B}(\partial A)=\eta \Box C & \delta_{B}B=0\\
 \end{array}
\end{equation}
where $\eta$ is an anticommuting space-time independent transformation parameter.
Particularizing for the 2 dimensional case the Lagrangian becomes 
\begin{equation}
 L_{B}=-\frac{1}{2}E^{2}+\bar{\psi}(i\gamma^{\mu}\partial_{\mu}-m)\psi-e\bar{\psi}\gamma^{\mu}A_{\mu}\psi+B(\partial A)+\frac{1}{2}B^{2}-i\partial_{\mu}\bar{C}\partial^{\mu}C
\end{equation}
and this can be rewritten after introducing another auxiliary field $\mathcal{B}$ as
\begin{equation}
 L_{\mathcal{B}}=\mathcal{B}E-\frac{1}{2}\mathcal{B}^{2}+\bar{\psi}(i\gamma^{\mu}\partial_{\mu}-m)\psi-e\bar{\psi}\gamma^{\mu}A_{\mu}\psi+B(\partial A)+\frac{1}{2}B^{2}-i\partial_{\mu}\bar{C}\partial^{\mu}C
\end{equation}
The dual BRST symmetry operators to be associated to the theory above in the 2 dimensional case are [29]
\begin{equation}
 \begin{array}{ll}
  \delta_{D}A_{\mu}=-\eta\epsilon_{\mu\nu}\partial_{\nu}\bar{C} & \delta_{D}\psi=-i\eta e \bar{C}\gamma_{5} \psi \\
\delta_{D}C=-i \eta\mathcal{B} & \delta_{D}\bar{C}=0\\
\delta_{D}\bar{\psi}=i\eta e \bar{C}\gamma_{5}\bar{\psi} & \delta_{D}F_{\mu\nu}=\eta\Box\bar{C}\\
\delta_{D}(\partial A)=0 & \delta_{D}B=0\\
\delta_{D}\mathcal{B}=0\\

\end{array}
\end{equation}
Moreover, as noted in reference [29] the interacting Lagrangian in 2 dimensions is invariant under the following transformations
\begin{equation}
 \begin{array}{ll}
C \rightarrow \pm i\gamma_{5}\bar{C} & \bar{C}\rightarrow \pm i \gamma_{5}C\\
\mathcal{B}\rightarrow \mp i\gamma_{5}B & A_{0}\rightarrow\pm i\gamma_{5}A_{1}\\
A_{1}\rightarrow\pm i \gamma_{5}A_{0} & B\rightarrow \mp i \gamma_{5}\mathcal{B}\\
E\rightarrow \pm i \gamma_{5}(\partial A) & (\partial A)\rightarrow \pm i \gamma_{5}E\\
e\rightarrow \mp i e & \psi\rightarrow\psi\\
\bar{\psi}\rightarrow\bar{\psi}\\
\end{array}
\end{equation}
Reference [29] shows that these are the analogues of the Hodge duality $(*)$ for this particular example and that they induce a discrete symmetry. 
One can also verify that
\begin{equation}
 * ( *\Phi )=\pm\Phi
\end{equation}
where for $(+)$ the generic field $\Phi$ is $\psi$, $\bar{\psi}$ and for $(-)$ $\Phi$ represents the rest of the fields. 
One can also observe that for the direct and dual BRST symmetries 
\begin{equation}
 \delta_{D}\Phi = \pm * \delta_{B} * \Phi
\end{equation}
is valid.
It has been known before that the above statements are valid for any even dimensional theory [27] and applications for $D=4$, $(3,1)$ and $D=6$ dimensional theories have been given. 
However, combining the ideas presented in section 2 with the observations in ref. [26] and some theorems of algebraic topology and geometry one can generalize the 
applicability of this method to any dimension. While it is true that in some cases non-local transformations emerge ([29]-[32]) the method described in this paper is simply 
a mathematical trick that allows the construction of dual theories with no sign problems so physical meaning of the artificial transformations is irrelevant.

\section{6. Atyiah-Singer index theorem and cohomology}
Up to this point I employed as a basic tool for the current construction, auxiliary fields of various types. These were introduced in order to construct a "theoretical measuring device" more suitable for theories presenting global anomalies and specifically for theories subject to an $SU(2)$ anomaly. 
I showed in the sections above that fictitious internal circular spaces may induce artificial discrete symmetries. These concepts have a direct analogy in the domain of (co)homology with torsion coefficient groups.
In categorial terms the connection between the construction using auxiliary fields and the construction using torsion coefficient groups can be written as the following commuting diagram
\begin{equation}
\begin{tikzcd}
\textfrak{F}_{\,S}^{\,n}(M) \arrow{r}{h}\arrow{d}{i} & \textfrak{F}_{\,S}^{\,2n}(M')\arrow{d}{j} \\
H^{p}(C,\mathbb{Z}) \arrow{r}{h^{*}}& H^{p}(C,\mathbb{G})
\end{tikzcd}
\end{equation}
Here, $\textfrak{F}_{\,S}^{\,n}(M)$ is the space of physical solutions of the theory containing an initial number $n$ of fields while 
$\textfrak{F}_{\,S}^{\,2n}(M')$ is the space of physical solutions for the theory obtained via the introduction of new auxiliary fields such that the required internal "circular" space emerges. This space contains the required topological particularity introduced via the employment of the auxiliary fields. 
It must be specified that the morphism in the lower arrow requires the use of the universal coefficient theorem. The upper arrow morphism is valid when we talk about the physical domain of the theory. $\mathbb{G}$ is a torsion group, e.g. $\mathbb{Z}_{p}$. If $Ext$ and/or $Tor$ are being taken into account in the construction of the respective spaces, the horizontal arrows become isomorphisms.

Simply stated, if we have a module $M$ over a ring $R$, an element $m\in M$ is called a torsion element of the module if there exists a regular element $r\in R$ (not a zero divisor) such that $r\circ m=0$. In the case of a group $G$, an element $g\in G$ is called a torsion element of the group if it has finite order i.e. if there is a positive integer $m$ such that $g^{m}=e$, $e$ being the $Id$ element of $G$. A (sub)-group is called torsion (sub)-group (or circular or periodic) if all its elements are torsion elements. Examples are $\mathbb{Z}_{p}$-groups with $p-$prime.

In this section I briefly introduce the Atyiah-Singer index theorem in the context of the Hodge-de-Rham theory. This is being done following mainly reference [35]. I also construct an analogy between the method presented above and the cohomology with coefficients in groups with torsion. By using the isomorphism between the de-Rham cohomology group and the group of harmonic functions ($\Delta\phi=0$, $\Delta=d\delta+\delta d$) over a closed manifold I introduce the general form for index theorems i.e. a connection between an index calculated in a topological respectively an analytical fashion. The presence of torsion in the coefficient structure of the cohomology, while preserving the isomorphism between cohomology and the group of harmonic functions (a result of the second Hodge theorem), also reflects the effect of an "internal circular space" as presented in the previous chapter from the perspective of index theorems. As I will show in what follows, different choices of coefficient groups may merge or dissociate classes in the cohomology groups. Due to the isomorphism with the group of harmonic forms, the same effect will be found on the analytic side of the index theorems. The universal coefficient theorems assure us that the choice of a rather unusual coefficient group has no physical effects if the $Ext$ and/or $Tor$ groups are correctly considered in the chain complex.
 For this I follow my previous result regarding the relativity of anomalies [34]. 
 For the sake of completeness I state here the following: 
 \par {\bf Theorem} (The Universal Coefficient Theorem) [34], [36]
\\ If $C$ is a chain complex of free abelian groups, then there are natural short exact sequences 
\begin{equation}
 0\rightarrow H_{n}(C)\otimes G\rightarrow H_{n}(C;G)\rightarrow Tor(H_{n-1}(C),G)\rightarrow 0
\end{equation}
$\forall$ $n$, $G$, and these sequences split. Here  $Tor(H_{n-1}(C),G)$ is the torsion group associated to the homology.
In this way homology with arbitrary coefficients can be described in terms of homology with the ``universal'' coefficient group $\mathbb{Z}$.
For cohomology the exact sequence changes into 
\begin{equation}
0\rightarrow Ext(H_{n-1}(C_{*}),G)\rightarrow H^{n}(C_{*};G)\rightarrow Hom(H_{n}(C_{*}),G)\rightarrow 0
\end{equation}
Here $Ext$ is the group extension. \\
$\flat$
\\
Relevant for the situation at hand is the following
\par{\bf Example} (Homotopy and coefficient group) [34], [36]
\\ Take a Moore space $M(\mathbb{Z}_{m},n)$ obtained from $S^{n}$ by attaching a cell $e^{n+1}$ by a map of degree $m$. 
The quotient map $f:X\rightarrow X/S^{n}=S^{n+1}$ induces trivial homomorphisms on the reduced homology with $\mathbb{Z}$ coefficients since the nonzero reduced homology groups of $X$ and $S^{n+1}$ occur in different dimensions. But with $\mathbb{Z}_{m}$ coefficients the situation changes, as we can see considering the long exact sequence of the pair $(X,S^{n})$, which contains the segment
\begin{equation}
 0=\tilde{H}_{n+1}(S^{n};\mathbb{Z}_{m})\rightarrow \tilde{H}_{n+1}(X;\mathbb{Z}_{m})\xrightarrow{f_{*}}\tilde{H}_{n+1}(X/S^{n};\mathbb{Z}_{m})
\end{equation}
Exactness requires that $f_{*}$ is injective, hence non-zero since $\tilde{H}_{n+1}(X;\mathbb{Z}_{m})$ is $\mathbb{Z}_{m}$, the cellular boundary map 
\begin{equation}
H_{n+1}(X^{n+1},X^{n};\mathbb{Z}_{m})\rightarrow H_{n}(X^{n},X^{n-1};\mathbb{Z}_{m})
\end{equation}
being exactly 
\begin{equation}
\mathbb{Z}_{m}\xrightarrow{m}\mathbb{Z}_{m}
\end{equation}
One can see that a map $f:X\rightarrow Y$ can have induced maps $f_{*}$ that are trivial for homology with $\mathbb{Z}$ coefficients but not so for 
homology with $\mathbb{Z}_{m}$ coefficients for suitably chosen $m$. This means that homology with $\mathbb{Z}_{m}$ coefficients can tell us that $f$ is not homotopic to a constant map, information that would remain invisible if one used only $\mathbb{Z}$-coefficients. 
$\flat$
\par This relatively simple example shows that torsion coefficient groups are in some sense "measuring devices" that allow us to consistently take into account global properties without inconsistencies.

\par Following ref. [33], it is precisely the $Tor/Ext$ correction that has an important effect on the presence of a sign ambiguity (an SU(2) anomaly). 
It is relevant at this point to remember that because on a compact Riemannian orientable manifold $M$, the cohomology group $H^{p}_{DR}(M)$ is isomorphic to the group of harmonic forms on $M$, $Harm^{p}(M)$, the two groups have the same dimension, hence
\begin{equation}
dim(H_{DR}^{p}(M))=dim(Harm^{p}(M))=b^{p}(M)
\end{equation}
where $b^{p}(M)$ is the Betti number of $M$. 
\par We can define the exterior derivative $d$ and the codifferential $\delta$ as adjoint to each other. On the ring of differential forms $\Lambda(M)$ on $M$ the action of $d$ induces a sequence 
\begin{equation}
0\rightarrow \Lambda^{0}(M)\xrightarrow{d_{0}}\Lambda^{1}(M)\xrightarrow{d_{1}}...\xrightarrow{d_{n-1}}\Lambda^{n}(M)\xrightarrow{d_{n}}0
\end{equation}
The codifferential generates another sequence of arrows oriented this time in the opposite direction
\begin{equation}
...\leftarrow \Lambda^{i-1}(M)\xleftarrow{\delta_{i-1}}\Lambda^{i}(M)\xleftarrow{\delta_{i}}\Lambda^{i+1}(M)\xleftarrow{\delta_{i+1}}...
\end{equation}
We now have 
\begin{equation}
\begin{array}{ccc}
(d_{i}\alpha,\beta)=(\alpha,\delta_{i}\beta), & \alpha\in\Lambda^{i}(M), &\beta\in\Lambda^{i+1}(M)
\end{array}
\end{equation}
Neither of these sequences is exact i.e. $Ker(d_{i})\neq Im(d_{i-1})$ and similarly for $\delta$. However $Im(d_{i-1})\subset Ker(d_{i})$ or equivalently $d^{2}=0$. This means that the first sequence is a de-Rham complex. To this complex we can associate the de-Rham cohomology groups which measure the lack of exactness of the sequence 
\begin{equation}
H_{DR}^{i}(M,R)=Ker(d_{i})/Im(d_{i-1})
\end{equation}
We define $\alpha\in\Lambda^{i}(M)$ to be co-closed (resp. co-exact) if $\alpha\in Ker(\delta_{i-1})$ (resp. $\alpha\in Im(\delta_{i}))$. We also can define the homogeneous Hodge-de-Rham operator $\Delta=(d+\delta)^{2}$. We then obtain on $\alpha\in\Lambda^{i}(M)$
\begin{equation}
\Delta_{i}=\delta_{i}d_{i}+d_{i-1}\delta_{i-1}
\end{equation}
Lets now assume that the $(i+1)$-form $\beta$ can be expressed as $\delta_{i+1}\beta'$, $\beta'\in\Lambda^{i+2}(M)$. If this is so, the product $(d_{i}\alpha,\delta_{i+1}\beta')$ is zero. Proceeding as in the previous section where the Hodge decomposition has been introduced, we conclude that
\begin{equation}
Im(d_{i-1})\,\bot\, Im(\delta_{i})\,\bot\, Ker(\Delta_{i})
\end{equation}
or, in other words that $\Lambda^{i}(M)$ has an unique splitting of the form 
\begin{equation}
\Lambda^{i}(M)=Im(d_{i-1})\oplus Im(\delta_{i})\oplus Ker(\Delta_{i})
\end{equation}
and consequently, since $Ker(\Delta_{i})=Harm^{i}(M)$ we have 
\begin{equation}
\alpha_{i}=d_{i-1}\alpha_{i-1}+\delta_{i}\alpha_{i+1}+h_{i},\; \alpha\in\Lambda^{i}(M)
\end{equation}
where $h_{i}$ is an harmonic $i$-form $\Delta_{i}h=0$. Every $i$-th de Rham cohomology class is represented by one and only one harmonic form 
\begin{equation}
H_{DR}^{i}(M,R)=Ker(\Delta_{i})=Ker(d_{i})/Im(d_{i-1})
\end{equation}
If the analytic index of the de Rham complex is now the integer defined by the alternating sum 
\begin{equation}
index(\Lambda(M),d)=\sum_{i}(-)^{i}dim(Ker(\Delta_{i}))
\end{equation}
we find 
\begin{equation}
index(\Lambda(M),d)=\sum_{i=0}^{N}(-)^{i}b^{i}(M)=\chi(M)=\int_{M}e(TM)
\end{equation}
Here, $e(TM)$ is the Euler class of the tangent bundle to $M$. 
The right-hand side of this expression is of a topological nature: it is a topological index. If $M$ is odd-dimensional, $index(\Lambda(M^{odd}),d)=0$ since $\chi(M^{odd})=0$. This remains true for index theorems of other differential operators. 
We may note that $\Delta$ and $d+\delta$ have the same Kernel (a harmonic form is closed and co-closed). Let us split 
\begin{equation}
\Lambda(M)=\Lambda^{even}(M)\oplus \Lambda^{odd}(M)
\end{equation}
in even and odd forms
 \begin{equation}
 \begin{array}{c}
\\
\Lambda^{even}(M)=\bigoplus_{i}\Lambda^{2i}(M)\\
\\
\Lambda^{odd}(M)=\bigoplus_{i}\Lambda^{2i+1}(M)\\
\\
\end{array}
\end{equation}
 Let $D_{+}$ and $D_{-}$ be the operators defined by 
\begin{equation}
\begin{array}{c}
\\
D_{+}=D=\sum_{i}(d_{2i}+\delta_{2i-1})\\
\\
D_{-}=D^{\dagger}=\sum_{i}(d_{2i-1}+\delta_{2i})\\
\\
\end{array}
\end{equation}
Then $D$ is a mapping 
\begin{equation}
D:\Lambda^{even}(M)\rightarrow \Lambda^{odd}(M)
\end{equation}
defined as
\begin{equation}
D(\alpha_{(0)},\alpha_{(2)},\alpha_{(4)},...)=(d_{0}\alpha_{(0)}+d_{1}\alpha_{(2)},d_{2}\alpha_{(2)}+d_{3}\alpha_{(4)},...)
\end{equation}
Its adjoint $D^{\dagger}$ is a mapping 
\begin{equation}
D^{\dagger}:\Lambda^{odd}(M)\rightarrow\Lambda^{even}(M)
\end{equation}
The associated Laplacians are given by 
\begin{equation}
\begin{array}{c}
\\
\Delta_{+}=D^{\dagger}D=\sum_{i}\Delta_{2i}\\
\\
\Delta_{-}=DD^{\dagger}=\sum_{i}\Delta_{2i-1}\\
\\
\end{array}
\end{equation}
Thus we can replace the definition of the analytical index of the de-Rham complex by 
\begin{equation}
index(\Lambda(M),D)=dim(Ker(\Delta_{+}))-dim(Ker(\Delta_{-}))
\end{equation}
or equivalently by 
\begin{equation}
index(\Lambda(M),D)=dim(Ker(D))-dim(Ker(D^{\dagger}))
\end{equation}
since 
\begin{equation}
\begin{array}{c}
\\
Ker(\Delta_{+})=Ker(D^{\dagger}D)=Ker(D)\\
\\
Ker(\Delta_{-})=Ker(DD^{\dagger})=Ker(D^{\dagger})\\
\\
\end{array}
\end{equation}

In fact, $Im(D)$ (resp. $Ker(D)$) is the orthogonal complement of $Ker(D^{\dagger})$ (resp. $Im(D^{\dagger})$). 
We also have that 
\begin{equation}
CoKer(D)=\Lambda^{odd}/Im(D)=Ker(D^{\dagger})
\end{equation}
and hence the analytic index may be given as
\begin{equation}
index(\Lambda(M),D)=dim(Ker(D))-dim(CoKer(D))
\end{equation}

The form of the analytic index for the de-Rham complex is not specific to this case. To see what it has in common with the index theorem for other complexes let us look at its general structure. 
First we notice that the de-Rham sequence should have been written 
\begin{equation}
0\rightarrow \Gamma(M,E_{0})\rightarrow\Gamma(M,E_{1})\rightarrow ... \rightarrow\Gamma(M,E_{i})\xrightarrow{D_{i}}\Gamma(M,E_{i+1})\rightarrow ... \rightarrow \Gamma(M,E_{n})\xrightarrow{D_{n}}0
\end{equation}
where $\Gamma(M,E_{i})$ is the module of cross sections of the vector bundle $E_{i}=\Lambda^{i}T^{*}M$ since the differential $i$-forms of $\Lambda^{i}(M)$ may be regarded as sections of the vector bundle $\Lambda^{i}T^{*}M$. The writing of the above sequence requires the existence of a differential operator of degree $1$ acting on a sequence of sections of vector bundles $E_{i}$ such that $D_{i+1}\circ D_{i}=0$. With this, the sequence qualifies as a complex. 
\par Secondly, the fact that the expression 
\begin{equation}
index(\Lambda(M),d)=\sum(-)^{i}dim(Ker(\Delta_{i}))
\end{equation}
was a well defined one was guaranteed by the nature of the Laplacian operator $\Delta=(d+\delta)^{2}$, the kernel of which is finite dimensional. This is a consequence of the fact that $\Delta$ is an elliptic operator.

 \par An elliptic operator defined on a compact manifold has a finite-dimensional kernel and cokernel and expressions of the type
 \begin{equation}
 index(\Lambda(M),D)=dim(Ker(D))-dim(CoKer(D))
 \end{equation}
 are well defined for them.

Looking at the de-Rham complex on the compact boundaryless manifold $M$ we conclude that it is an elliptic complex since its associated Laplacians are elliptic. 
 Let $E$ be a vector bundle. An elliptic complex $(E,D)$ is a finite sequence of differential operators $D_{i}:\Gamma(M,E_{i})\rightarrow \Gamma(M,E_{i+1})$ acting on smooth sections such that $D_{i+1}\circ D_{i}=0$ and the Laplacians of the complex $\Delta_{i}=D_{i}^{\dagger}D_{i}+D_{i-1}D_{i-1}^{\dagger}$ where $D_{i}^{\dagger}$ is the adjoint operator with respect to the scalar product on the fibres with a smooth density on $M$, are elliptic on $\Gamma(M,E_{i})$. 

 \par Since $(D_{i+1}\circ D_{i})^{\dagger}=D_{i}^{\dagger}\circ D_{i+1}^{\dagger}$, it follows that if the complex $(\Gamma(M,E_{i}),D_{i})$ is elliptic, so is the complex $(\Gamma(M,E_{i+1}),D_{i}^{+})$ where the arrows point in the opposite direction. 
 To relate this picture to the form of the index for a de Rham complex we have to reduce the elliptic complex to a two-term elliptic complex (to roll up the complex) and see that the new complex has the same index as the original one $(\Gamma(M,E),D)$. This is where the comparison with the previous equations for the index comes in. Defining the even and odd bundles $E^{even}=\bigoplus_{i}E_{2i}$, $E^{odd}=\bigoplus_{i}E_{2i+1}$
 \begin{equation}
 \begin{array}{cc}
 \Gamma(M,E^{even})=\bigoplus_{i}\Gamma(M,E_{2i}), & \Gamma(M,E^{odd})=\bigoplus_{i}\Gamma(M,E_{2i+1})\\
 \\
 D=\bigoplus_{i}(D_{2i}+D_{2i-1}^{\dagger}), & D^{\dagger}=\bigoplus_{i}(D_{2i-1}+D_{2i}^{\dagger})\\
 \\
 \end{array}
 \end{equation}
 and the associated Laplacian 
\begin{equation}
\begin{array}{c}
\\
\Delta_{i}=D_{i}^{\dagger}D_{i}+D_{i-1}D_{i-1}^{\dagger}\\
\\
\Delta_{+}=\sum_{i}\Delta_{2i}=D^{\dagger}D\\
\\
\Delta_{-}=\sum_{i}\Delta_{2i-1}=DD^{\dagger}\\
\\
\end{array}
\end{equation}
 The analytical index of an elliptic complex $(\Gamma(M,E),D)$ is defined to be the integer
 \begin{equation}
 index(\Gamma(M,E),D)=\sum_{i}(-)^{i}dim(Ker(\Delta_{i}))=dim(Ker(\Delta_{+}))-dim(Ker(\Delta_{-}))
 \end{equation}

 \par We note that the differential operator defining a complex, the Riemannian scalar product defining its adjoint and the ellipticity property which guarantees that the rhs of the equation above is well defined (an integer) are the ingredients for the definition of an index of a compact manifold. In order to have a non-trivial index the operator $D$ cannot be self-adjoint. The Atiyah-Singer index theorem states that the analytic index is equal to the topological index of the complex, which is given by the rhs in the formula of the Atiyah-Singer index theorem. The statement of this theorem is as follows: 
 \par Let $(\Gamma(M,E),D)$ be an elliptic complex over a compact boundaryless manifold $M$ of even dimension $n$. Then the index of the complex is given by 
 \begin{equation}
 index(\Gamma(M,E),D)=(-)^{n(n+1)/2}\int_{M}ch(\bigoplus_{i=0}^{n}(-)^{i}E_{i})\frac{Td(TM^{C})}{e(TM)}
 \end{equation}
 $Td(TM^{C})$ is the Todd class of the complexified tangent bundle $TM^{C}$ and $e(TM)$ is the Euler class. 
In the above integrand only $n$-forms are retained. If the manifold is odd-dimensional, the index of the differential operator $D$ is zero. Due to this trivial situation it makes sense to go to an even-dimensional field-space as explained in the BRST-dual-BRST construction of the previous sections.

 \par At this moment we can see how the main construction of this article affects the Atiyah-Singer theorem. The next splitting due to the Kahler structure introduced over the field space gives
\begin{equation}
\begin{array}{ll}
 T_{(1,0)}=\{v\in T_{x}\mathcal{M}^{\mathbb{C}}|J_{x}(v)=iv\}; &  T_{(0,1)}=\{v\in T_{x}\mathcal{M}^{\mathbb{C}}|J_{x}(v)=-iv\} \\
\end{array}
\end{equation}
The isomorphism between the cohomology group and the group of harmonic forms is preserved. The introduction of a Kahler structure over the field space has two effects. First it allows the construction of an explicit internal circular space and second, it allows a different splitting, one that dissociates the two different signs that can arise when a large gauge transformation is performed. We can repeat the same discussion as above, only this time with a non-trivial coefficient structure in cohomology. In fact one can chose the torsion of the coefficient groups in cohomology such that they compensate precisely the topological properties of the $SU(2)$ group. If, for example, the coefficient group in cohomology is $\mathbb{Z}_{2}$ the two regions of positive and negative eigenvalues that make the path integral associated to the $SU(2)$ problem inconsistent become properly separated. The coefficient group in cohomology now contains different classes. 
The isomorphism between the cohomology group and the group of harmonic forms on $M$ was until now understood as
\begin{equation}
H^{p}_{DR}(M;\mathbb{Z}) \cong Harm^{p}(M;\mathbb{Z})
\end{equation}
The universal coefficient theorem assures us that we can use a different coefficient group. One choice then is
\begin{equation}
H^{p}_{DR}(M;\mathbb{Z}_{2}) \cong Harm^{p}(M;\mathbb{Z}_{2})
\end{equation}
This choice can be used such that the distinction between the two regions of different signs is made explicit. 
Let's now take the dimension of the above construction. As the isomorphism is preserved the dimensions of the two groups will be the same, albeit different from the case above. Indeed 
\begin{equation}
dim(H_{DR}^{p}(M;\mathbb{Z}_{2}))=dim(Harm^{p}(M;\mathbb{Z}_{2}))=b^{p}(M)_{\mathbb{Z}_{2}}
\end{equation}
The introduction of inner space circular integration paths is translated in coefficient groups in cohomology. This leads to a reorganization of the integration such that, simply stated, 
\begin{equation}
\underbrace{\{i_1,i_2,...,i_n\}}_{\pm 1}\rightarrow\{\underbrace{(i_1,...,i_q)}_{-1},\underbrace{(i_{q+1},...,i_n)}_{+1}\}
\end{equation}
where $i_{p}$ represent points on the non-trivial manifold where the integration is performed in the two cases (with trivial coefficient group and with $Z_{2}$ coefficient group). 
If the first subset on the right is characterized by a positive sign and the second by a negative sign then the specific choice of a torsional (periodic) coefficient group in cohomology makes the two domains clearly separated and well indexed. 
I explained in the introduction of this article the origin of the Bose-Einstein and Fermi-Dirac statistics as a result of how the topology of the quotient space $X/S_{n}$ where $S_{n}$ is the symmetry group, changes with respect to the original space $X$. While the two spaces remain isomorphic, the global properties differ according to the number of dimensions considered. It is interesting to see how it is possible to relate the case with $dim(X)=1,2$ to the case $dim(X)\geq 3$. Indeed, in dimensions larger than $2$ performing two rotations around a singularity brings us to a curve that can be homotopically deformed into a point. However, the integration is sensible to homology and cohomology. In principle the homology groups $H_{k}(C)$ of a chain complex $C$ relate to the shape of the manifold. The cohomology groups $H^{k}(C)$ relate to the differential forms defined over a manifold. Hence if we have a manifold $M$ characterized by a sequence of homology groups then one can define the integral 
\begin{equation}
\int_{M}\omega
\end{equation}
as being characterized by the differential form $\omega$ and by the manifold $M$. Integration can be seen as the pairing 
\begin{equation}
H_{k}(M,\mathbb{R})\times H^{k}(M,\mathbb{R})\rightarrow \mathbb{R}
\end{equation}
such that 
\begin{equation}
([M],[\omega])\rightarrow \int_{M}\omega
\end{equation}
where this pairing is constructed with real coefficients and this coefficient structure characterizes also the measure of integration and implicitly the differential form $\omega$. Here, $[M]$ represents a class in homology and $[\omega]$ represents a class in cohomology. The pairing above is an isomorphism only when this particular choice of coefficients is made. For other coefficients this pairing may fail to be an isomorphism, the correction being controlled by the $Ext$ and $Tor$ groups. The pairing then becomes 
\begin{equation}
H_{k}(M,\mathbb{G})\times H^{k}(M,\mathbb{G})\rightarrow \mathbb{G}
\end{equation}
The same principle translates for functional integration in the partition function. Because of this, the case for $dim\geq 3$ is anomalous only if certain unsuitable choices of coefficient groups in (co)homology are made. Otherwise, the integration (which is seen as a pairing between (co)homology) is itself defined via a torsion coefficient group which acts as an "anti-anomaly". The physical aspects of the original theory are however preserved, in the $Tor$ group, albeit not in an explicit way.

\section{7. conclusion}
This paper proposes a potential theoretical tool capable to explain why hedgehog structures may be found experimentally. It also opens new perspectives on quantum computing via deconfined 
hedgehogs obeying non-abelian statistics. Although this paper does not solve all problems related to the practical construction of topological quantum computers it makes the concept theoretically plausible. 
On the theoretical side, the conclusion of this article is that global, topological anomalies are a reflection of the fact that unsuitable topological "measuring tools" are being used. These "tools" are analogous to the coefficient groups in (co)homology. The choice of those coefficient groups is arbitrary in the sense that universal coefficient theorems relate (co)homologies with various coefficient structures. Some of these choices can make the theories well defined over non-trivial topologies.

\section{Appendix}
\section*{A. Kahler Manifolds and the field-anti-field formalism}
The main ideas of this paper (symmetry out of cohomology and dual gauge fixing) define a new way in which symmetry can be regarded. Instead of regarding symmetry as given by nature, here, some discrete symmetries are considered as artificial tools that can be added and removed from the theory. In order to make this clear I used the field-antifield formalism, a mathematical construction that relies on the Batalin-Vilkovisky quantization prescription.
\par This is a method that has been widely used in quantum gravity and string (field) theory. Nevertheless, this work does not rely on any string theory or quantum gravity assumptions and is completely self consistent in the context of gauge theories and quantum field theories (although new applications to string theory are not excluded). Essentially any theory can be extended by following the field-anti-field prescription. The resulting theory, equivalent to the previous one (dual) can be constructed in such a way that a Kahler structure becomes manifest[20]. 
\par As has been shown in [21] the field-antifield and the antibracket formalisms have a geometrical interpretation. The Batalin Vilkovisky formalism has also been set up for curved supermanifolds of fields and anti-fields with a fermionic symplectic structure [22]. Once a Kahler structure is introduced the symplectic structure is reduced to that given by a fermionic Kahler 2-form[21], [20]. 
\par The specific way in which the new structure is induced is by introducing a set of auxiliary fields that can be seen as shifts in the field space. After performing two shifts one obtains a BRST-anti-BRST structure constructed in a way that enforces the Schwinger Dyson equations as Ward identities. In general the Schwinger Dyson equations are the quantum equations of motion. They are derived as a consequence of the generalization to path integrals of the invariance of an integral under a redefinition of the integration variable from $x$ to $x+a$. The BRST-anti-BRST symmetry was used in order to enforce precisely this at the level of Ward identities. The dual symmetry is obtained analogously by using an internal space. Precisely this method of finite shifting in the field space ensures that no divergencies in any of the kernel momenta appear due to this procedure. In fact the resulting object can be regarded as being shifted (in some directions defined for some artificial well behaving internal spaces) and finite shifts are not expected to alter the momenta of the kernel (variance, etc.)
\par It is also important to ensure that the field transformations provide the required form for the Jacobian. This is clear from the way in which the field structure is constructed: auxiliary fields are introduced in the sense of the field-anti-field formalism in pairs such that the overall field space becomes Kahlerian. As will be shown in the next section of this supplemental material, the Kahlerian structure is by definition one that assures a time-reversal type symmetry on the field structure and on the Jacobian and this structure is encoded in the field-anti-field formalism. 
\par Of course, the discrete symmetry emerges after one performs two transformations with the ultimate goal of obtaining a BRST-anti-BRST symmetry together with the associated dual symmetry. 
One can also ask if it is possible to perform other transformations that change the Jacobian in a different way. The answer is of course yes, but the final symmetry must be obtained for the entire structure i.e. the action and the integration measure. Performing the transformation as specified in section 2 and compensating every time for the transformations of the measure will produce the same Kahler structure and the same "time-reversal-type" symmetry which will result in the same global symmetry for the resulting determinant [20]. 
\par In order to be more specific let me focus on a general example. Let $[dq]$ be my initial measure, $G_{a}$ a transformation of the fields and $S[q]$ be my action. $[dq]$ is assumed not to be invariant under $G_{a}$. By construction $S[q]$ is considered invariant and so will also be $S'[q',a]$ where $a$ is the parameter of the transformation. One assumes the integration over $a$ as being trivial. Performing the change in variables $q\rightarrow q'$ will affect $[dq]$. The resulting transformation will be
\begin{equation}
\int [dq]\rightarrow \int [dq']det\lvert \frac{\partial q_{i}}{\partial q'_{j}}\rvert=\int [dq']det(M_{ij})
\end{equation}
\par Here the measure $[dq']$ is not invariant under the gauge transformations. The determinant of the transformation is also not invariant but the invariance is recovered when one combines the two transformations. Then, the gauge fixing procedure can be performed and one obtains the emerging global (anti)BRST symmetry. Please note that at this level the Jacobian has no special discrete symmetry. On the dual "branch" one can do the same thing obtaining the dual(anti)BRST symmetry. Only after generating the internal space over which one defines the dual BRST symmetry one can define the hodge star operation which induces a discrete time reversal type symmetry over the entire field space and implicitly over the resulting block-determinant.
\par In order to improve on clarity let's think in the terms of the field-anti-field formalism. For the sake of simplicity the field space can be regarded as a $D$ dimensional manifold parametrized by real coordinates $y^{i}=(y^{1},y^{2},...,y^{D})$. After performing the field extension in the sense of Batalin-Vilkovisky the space is extended to a $2D$ dimensional manifold of the form $y^{i}=(x^{1},x^{2},...,x^{D},\xi^{1},\xi^{2},...,\xi^{D})$ where $x$ are the bosonic and $\xi$ are the fermionic coordinates. Now the space has a symplectic structure given by a closed non-degenerate 2-form
\begin{equation}
\omega=dy^{j}\wedge dy^{i}\omega_{ij}
\end{equation}
\begin{equation}
d\omega = 0
\end{equation}
Finally an antibracket structure emerges 
\begin{equation}
\{A,B\}=A\overleftarrow{\partial}_{i}\omega^{ij}\partial_{j}B
\end{equation}
By introducing the internal space in the way explained in section 2 one extends the space again. Now $D=2d$ and a hodge star operation (and its associated duality) becomes well defined. Having the Kahler structure defined by the tensor 
\begin{equation}
J = \left( \begin{array}{cccc} 0 & 1 & 0 & 0\\ -1 & 0 & 0 & 0\\0 & 0 & 0 & 1\\ 0 & 0 & -1 & 0 \end{array}\right)
\end{equation}
and going to a complex coordinate basis 
\begin{equation}
\begin{array}{ccc}
z^{a}=(z^{\alpha},\zeta^{\alpha}) & \bar{z}^{a}=(\bar{z}^{\alpha},\bar{\zeta}^{\alpha}),&\alpha = 1,2,...,d
\end{array}
\end{equation}
\begin{equation}
\begin{array}{cc}
z^{\alpha}=x^{\alpha}+ix^{d+\alpha} & \zeta^{\alpha}=\xi^{\alpha}+i \xi^{d+\alpha}\\
\end{array}
\end{equation}
we obtain a supermanifold with a Kahlerian geometry and an equivalent change in the representation of the antibracket. 
Following reference [23] (for the sake of brevity I will not perform the calculations here again) the change in the metric which amounts to the redefinition of the poisson bracket (generalized to the antibracket in our situation)
\begin{equation}
\{f,g\}=\sum_{\alpha\beta}\Omega^{\alpha,\beta}\frac{\partial f}{\partial\eta^{\alpha}}\frac{\partial g}{\partial \eta^{\beta}}
\end{equation}
modifies the expression of the integration measure taking the change of the metric in the definition of the antibracket and mapping it onto the structure of the resulting global block-determinant. (see eq. (11)-(15) and (17)-(18) of ref. [23]). This ensures that the discrete symmetry affects the resulting determinant in the desired way. 
\par Another way of looking at this discrete symmetry is to consider it as induced by the antipode of a hopf-algebra (the vector space analogue of the Hodge star). Only after one constructs the global BRST-anti-BRST and dual-BRST-anti-BRST symmetries will the discrete symmetry emerge and the method of constructing the first two symmetries already implies the inclusion of the Jacobian of the considered transformations in obtaining the final symmetries involving the action as well as the measure of integration (see [18],[19]).
\par As an interlude, one may observe that here, I used the cohomology and Hodge duality in order to generate a discrete symmetry. Further symmetries could be obtained considering other topological properties like cobordism or Morse-surgery.

\section*{B. Mathematical aspects of Kahler manifolds}
This section is a short review of some relevant aspects related to the Kahler manifolds. More general discussions can be found in [24], [20] and [13].
Having a differential manifold $M$ and a tensor of type $(1,1)$ $J$ such that  $\forall p \in M$, $J_{p}^{2}=-1$, the tensor $J$ will give a structure to $M$ with the property that the eigenvalues of 
it will be of the form $\pm i$.
This means that
 $J_{p}$ is an even dimensional matrix and $M$ is an even manifold. From the same definition it follows that $J_{p}$ can divide the complexified tangent space at $p$ in two disjoint vector subspaces
\begin{equation}
 T_{p}M^{C}=T_{p}M^{+}\oplus T_{p}M^{-}
\end{equation}
\begin{equation}
 T_{p}M^{\pm}=\{Z\in T_{p}M^{C}|J_{p}Z=\pm iZ\}
\end{equation}
One can introduce two projection operators of the form
\begin{equation}
 P^{\pm}:T_{p}M^{C}\rightarrow T_{p}M^{\pm}
\end{equation}

\begin{equation}
 P^{\pm}=\frac{1}{2}(1\pm iJ_{p})
\end{equation}
which will decompose Z as $Z=Z^{+}+Z^{-}$. This construction will generate a holomorphic and an antiholomorphic sector: $Z^{\pm}=P^{\pm}Z\in T_{p}M^{\pm}$, $T_{p}M^{+}$ being the holomorphic sector.
A complex manifold appears when demanding that given two intersecting charts $(U_{i},\gamma_{i})$ and $(U_{j},\gamma_{j})$, the map $\psi_{ij}=\gamma_{j}\gamma_{i}^{-1}$ from $\gamma_{i}(U_{i}\cap U_{j})$
 to $\gamma_{i}(U_{i}\cap U_{j})$ is holomorphic. Here $\gamma_{i}$ and $\gamma_{j}$ are chart homeomorphisms and $\psi_{ij}$ is the transition map. In this case the complex structure is given independently from the chart by \begin{equation}
 J_{p} = \left( \begin{array}{cc} 0 & 1 \\ -1 & 0 \end{array} \right) \forall p\in M
\end{equation}
In the complex case there is a unique chart-independent decomposition in holomorphic and antiholomorphic parts. This means we can now choose as a local basis for those subspaces the vector
$(\frac{\partial}{\partial z^{\mu}},\frac{\partial}{\partial\bar{z}^{\mu}})$ where ($z^{\mu},\bar{z}^{\mu}$) are the complex coordinates
such that the complex structure becomes
\begin{equation}
 J_{p} = \left( \begin{array}{cc} i1 & 0 \\ 0 & -i1 \end{array} \right) \forall p\in M
\end{equation}
If we add a Riemannian metric $g$ to the complex manifold and demand that the metric satisfies $g_{p}(J_{p}X,J_{p}Y)=g_{p}(X,Y), \forall p\in M$ and $X,Y\in T_{p}M$ then the metric is called 
hermitian and $M$ is called a hermitian manifold. A complex manifold always admits a hermitian metric. 
Using the base vectors of the complexified $T_{p}M^{C}$ we can always write the metric locally as
\begin{equation}
 g=g_{\mu\bar{\nu}}dz^{\mu}\otimes d\bar{z}^{\nu} +g_{\bar{\mu}\nu}d\bar{z}^{\mu}\otimes dz^{\nu}
\end{equation}
If we have a hermitian manifold $(M,g)$ with $g$ Hermitian metric and a fundamental 2-tensor $\Omega$ whose action on vectors $X$ and $Y\in T_{p}M$ is 
\begin{equation}
 \Omega_{p}(X,Y)=g_{p}(J_{p}X,Y)
\end{equation}
then we call $\Omega_{p}(X,Y)$ a Kahler form. 
With this definition the Kahler form has some very useful properties. 
Firstly it is antisymmetric
\begin{equation}
 \Omega(X,Y)=g(J^{2}X,JY)=-g(X,JY)=-\Omega(Y,X)
\end{equation}
Then it is invariant under the action of the complex structure
\begin{equation}
 \Omega(JX,JY)=\Omega(X,Y)
\end{equation}
and under complexification 
\begin{equation}
 \Omega_{\mu\nu}=i g_{\mu\nu}=0
\end{equation}
\begin{equation}
 \Omega_{\bar{\mu}\bar{\nu}}=i g_{\bar{\mu}\bar{\nu}}=0
\end{equation}
\begin{equation}
 \Omega_{\mu\bar{\nu}}=-\Omega_{\bar{\nu}\mu}=ig_{\mu\bar{\nu}}
\end{equation}
thus leading to
\begin{equation}
 \Omega=ig_{\mu\bar{\nu}}dz^{\mu}\wedge d\bar{z}^{\nu}
\end{equation}
A Kahler manifold is a hermitian manifold $(M,g)$ whose Kahler form $\Omega$ is closed ($d\Omega$=0). $g$ is called a Kahler metric.
The closing condition defines a differential equation for the metric. 
\begin{equation}
 d\Omega = (\delta+\bar{\delta})ig_{\mu\bar{\nu}}dz^{\mu}\wedge d\bar{z}^{\nu}=
\end{equation}
\begin{equation}
\frac{i}{2}(\delta_{\lambda}g_{\mu\bar{\nu}}dz^{\lambda}\wedge dz^{\mu}\wedge d\bar{z}^{\nu})+\frac{i}{2}(\delta_{\bar{\lambda}}g_{\mu\bar{\nu}}-\delta_{\bar{\nu}}g_{\mu\bar{\lambda}})d\bar{z}^{\lambda}\wedge dz^{\mu}\wedge d\bar{z}^{\nu}=0
\end{equation}
This leads to the relations
\begin{equation}
 \frac{\delta g_{\mu\bar{\nu}}}{\delta z^{\lambda}}=\frac{\delta g_{\lambda\bar{\nu}}}{\delta z^{\mu}}
\end{equation}
\begin{equation}
 \frac{\delta g_{\mu\bar{\nu}}}{\delta \bar{z}^{\lambda}}=\frac{\delta g_{\mu\bar{\lambda}}}{\delta \bar{z}^{\nu}}
\end{equation}
The solution of the above equation takes the form $g_{\mu\bar{\nu}}=\delta\bar{\delta}K_{i}$ on a chart $U_{i}$ included in the manifold $M$.
$K_{i}$ is called Kahler potential.

\begin{equation}
 K_{i}:U_{i}\rightarrow R
\end{equation}
\begin{equation}
 K_{i}=K_{i}^{*}
\end{equation}
The Kahler form can be locally expressed in terms of the Kahler potential as
\begin{equation}
 \Omega=i\delta\bar{\delta}K_{i}
\end{equation}
The definition given above is the most general one. This can of course be extended to the field space of the problem analyzed in the section 2. The procedure explained there generates the (dual)field-anti-field structure required to make the link with the Kahler structure described above. 
\par I will continue by reviewing some further mathematical concepts: 
\paragraph{Hodge-* operator}
Let $(M,g)$ be a Riemannian 4-manifold for which we can define the * operator in the following way [25]:
\begin{equation}
 \alpha \wedge * \beta =g_{p}(\alpha,\beta)dv_{g}
\end{equation}
\begin{equation}
  \alpha,\beta \in \bigwedge^{2}
\end{equation}
We have also that $(**)=1$ on $\bigwedge^{2}$ which means that $\bigwedge^{2}$ splits into eigenspaces as 
\begin{equation}
 \bigwedge^{2}=\bigwedge_{+}^{2}+\bigwedge_{-}^{2}
\end{equation}
where the two eigenspaces correspond to eigenvalues +1 and -1 respectively. A 2 form which belongs to $\bigwedge_{+}^{2}$ is called self-dual whereas if it belongs to the other eigenspace it is called 
anti-self-dual. An important remark to be done here is that given a p-vector $\lambda\in\bigwedge^{p}V$ then $\forall \theta\in\bigwedge^{n-p}V$ there exists the wedge product such that $\lambda\wedge\theta\in\bigwedge^{n}$.
\paragraph{Hodge Theorem}
Let me define the following 3 operators
\begin{equation}
 d:C^{k}\rightarrow C^{k+1}
\end{equation}
\begin{equation}
 d^{*}=*d*:C^{k}\rightarrow C^{k-1}
\end{equation}
\begin{equation}
 \Delta=d d^{*}+d^{*} d:C^{k}\rightarrow C^{k}
\end{equation}
as being in order the exterior differential, the coexterior differential and the Laplace operator. 
The exact and co-exact forms are orthogonal. The Hodge theorem allows the identification of a unique representative for each cohomology class as belonging to the Kernel of the Laplacian defined for the specific complex manifold. 
If this is put together with the definition of the Kahler manifold we obtain extra symmetries in the Hodge structure of the manifold. 
\par As noted in reference [21] and [20]  the field-anti-field structure is amenable to the construction of a Kahlerian structure imposed on the system of fields. 
If one thinks at the antipode in a Hopf algebra one can see that there are not few similarities between the hodge star operator and the antipode. Indeed, the hodge star induces a symmetry that can be 
identified with time-reversal in the case of Kahlerian structures. All one has to do is to suitably introduce fields and antifields via appropriate trivial symmetries such that the antipodal structure becomes visible.

\section*{C. BRST-anti-BRST}
\par The BRST quantization and the gauge fixing procedure can be seen together as a canonical transformation acting on the field structure of the theory. The method presented in the main article that allowed the 
construction of internal spaces and the definition of Hodge-dual operations can be used to make the time reversal type symmetry manifest in any theory. 
\par What one must consider is the full de-Rham cohomology and identify the operators of BRST with the de-Rham cohomology operators. In order to do this one observes that the standard BRST-anti-BRST structure is not sufficient. In fact there 
exists another structure called the dual (anti)-BRST. 
\par This structure is the analogue of the co-exterior derivative of differential geometry in the way in which the anticommuting (anti)-BRST transformations are the analogue of the exterior derivative. Imposing the BRST-anti-BRST symmetries together with the dual-BRST-dual-anti-BRST symmetries via a collective field approach results in a theory cohomologically equivalent with the original one that contains an extra discrete symmetry. This symmetry can be used in order to fix the positive definiteness of the fermionic determinant and to solve the sign problem of the $SU(2)$ anomaly.
\par The presence of the new collective fields allow for extensions of the BRST symmetry. These can be used (while keeping the gauge fixing) in some innovative ways. In practice any extension of the field-antifield structure is allowed. The only condition is that the resulting extended action satisfies the master equation
\begin{equation}
 (S,S)=0
\end{equation}
The antibracket used above is just a generalization of the Poisson structure for field-anti-field extended actions. It is defined as
\begin{equation}
 (F,G)=\frac{\delta^{R}F}{\delta\phi^{A}}\frac{\delta^{L}G}{\delta\phi^{*}_{A}}-\frac{\delta^{R}F}{\delta\phi^{*}_{A}}\frac{\delta^{L}G}{\delta\phi^{A}}
\end{equation}
The BRST transformation in the extended case can be seen as given by the antibracket where the generator of the transformation is the generalized action
\begin{equation}
 \delta F = (S,F)
\end{equation}
The nilpotency of the BRST transformation is reflected in the condition
\begin{equation}
 (S,S)=0
\end{equation}
which is called the ``classical master equation''. "Quantum" corrections to this formula may appear in some cases mainly when integration is performed over only one (fermionic) field in the field-antifield pair.
 These amount to changes in the BRST transformation rules. A good explanation of the interplay between "quantum" and "classical" master equations on one side and the transformation rules and choices of field
 structures on the other side can be found in [18], [19].

\end{document}